\documentclass[preprint2]{aastex}

\newcommand{\zabs}{$z_{\rm abs}\,$}
\newcommand{\hh}{H$_2\,\,$}

\slugcomment{Accepted to ApJ}

\shorttitle{H$_2$, D/H and metals at \zabs = 3.025 
toward Q~0347--3819}
\shortauthors{Levshakov et al.}

\def\la{\;
\raise0.3ex\hbox{$<$\kern-0.75em\raise-1.1ex\hbox{$\sim$}}\; }
\def\ga{\;
\raise0.3ex\hbox{$>$\kern-0.75em\raise-1.1ex\hbox{$\sim$}}\; }

\def\la{\;
\raise0.3ex\hbox{$<$\kern-0.75em\raise-1.1ex\hbox{$\sim$}}\; }
\def\ga{\;
\raise0.3ex\hbox{$>$\kern-0.75em\raise-1.1ex\hbox{$\sim$}}\; }

\begin{document}
\title{Molecular hydrogen, deuterium and metal abundances
in the damped Ly$\alpha$ system
at \zabs = 3.025 toward QSO 0347--3819
\altaffilmark{1}
}
\author{Sergei A. Levshakov\altaffilmark{2,3},
Miroslava Dessauges-Zavadsky\altaffilmark{4,5}, 
Sandro D'Odorico\altaffilmark{4},
}
\and
\author{
Paolo Molaro\altaffilmark{6}
}
\altaffiltext{1}
{Based on public data released from
UVES Commissioning at the VLT Kueyen telescope, ESO, Paranal,
Chile. }
\altaffiltext{2}{Division of Theoretical Astrophysics,
National Astronomical Observatory, Tokyo 181-8588, Japan;
lev@th.nao.ac.jp.}
\altaffiltext{3}{Department of Theoretical Astrophysics,
Ioffe Physico-Technical Institute, St. Petersburg 194021, Russia;
lev@astro.ioffe.rssi.ru.}
\altaffiltext{4}{European Southern Observatory, 
Karl-Schwarzschild-Str. 2,
D-85748 Garching bei M\"unchen, Germany;
sdodoric@eso.org.}
\altaffiltext{5}{Observatoire de Gen\`eve, CH--1290 Sauverny,
Switzerland; mdessaug@eso.org.}
\altaffiltext{6}{Osservatorio Astronomico di Trieste, 
Via G.B. Tiepolo 11, I-34131 Trieste, Italy; molaro@oat.ts.astro.it.}

\begin{abstract}
We have detected 
in high resolution spectra of the quasar Q0347--3819
obtained with the UVES
spectrograph at the VLT/Kueyen telescope 
over 80 absorption features in the Lyman and
Werner \hh bands at the redshift of
a damped Ly$\alpha$ system 
at \zabs = 3.025. 
At the same redshift, numerous absorption lines of atoms
and low ions (\ion{H}{1}, \ion{D}{1}, \ion{C}{2},
\ion{C}{2}$^\ast$, \ion{N}{1}, \ion{O}{1}, \ion{Si}{2},
\ion{P}{2}, \ion{Ar}{1}, \ion{Cr}{2}, \ion{Fe}{2},
and \ion{Zn}{2}) were identified.
The \zabs = 3.025 system spans over $\sim 80$ km~s$^{-1}$
and exhibits a multicomponent velocity structure
in the metal lines.
The main component at \zabs = 3.024855 shows a
total \hh column density  $N({\rm H}_2) =
(4.10\pm0.21)\times10^{14}$ cm$^{-2}$ and
a fractional molecular abundance
$f_{{\rm H}_2} = (1.94\pm0.10)\times10^{-6}$
derived from the \hh lines
arising from J = 0 to 5 rotational levels of the ground
electronic-vibrational state. 
In the second strong component 
at \zabs = 3.024684
we were able to set 
a stringent upper limit to $f_{{\rm H}_2}$  of
$3.3\times10^{-9}$ ($3\sigma$ c.l.).
The population of the J levels can be represented by a single
excitation temperature of $T_{\rm ex} = 825\pm110$~K.
The ortho:para-\hh ratio equals $3.1\pm0.3$. This ratio
is comparable to that calculated from hot \hh formation 
upon grain surfaces (ortho:para-\hh = 3:1), but is higher
than the `freeze-out' quantity (ortho:para-\hh = 1:4) 
predicted for cosmological epochs at $z < 20$, 
implying recent formation of the hydrogen molecules
following photo-dissociation of pristine \hh.
In the main component the properties of the neutral atoms
and low ions can be described with the same 
broadening parameter $b \simeq 3$ km~s$^{-1}$ of the
\hh cloud.
The rate at which UV photons are absorbed in the Lyman
and Werner bands is found to be
$\beta_0 \simeq 2\times10^{-9}$ s$^{-1}$.
Taking into account that similar $\beta_0$ values are observed in the
Galactic diffuse molecular clouds,
it shows that the UV radiation fields in the DLA and in the MW
are very much alike.
We also found the \hh formation rate of $R\,n_{\rm H} 
\simeq 3\times10^{-16}$ s$^{-1}$  at \zabs = 3.025 which is consistent with
observations in the Galactic disk diffuse clouds 
where $n_{{\rm H}_2}$ is low.
Comparing the relative rotational population ratios of \hh
with the measured \ion{C}{2}$^\ast$/\ion{C}{2} ratio,
we can infer that  $n_{\rm H} \simeq 6$ cm$^{-3}$
in the main component, which gives a size 
of $D \simeq 14$ pc along the line of sight.
The mean value of the dust-to-gas ratio 
estimated for the whole \zabs = 3.025 system
from the [Cr/Zn] ratio
is equal to $\tilde{k} = 0.032\pm0.005$
(in units of the mean Galactic ISM value).
For the first time we unambiguously reveal a pronounced
[$\alpha$-element/iron-peak] enhancement of
[O,Si/Zn] = $0.6\pm0.1$ (6$\sigma$ c.l.) at high redshift.
The simultaneous analysis of metal and hydrogen lines leads to
D/H = $(3.75\pm0.25)\times10^{-5}$.
This value is consistent with standard big bang 
nucleosynthesis if the baryon-to-photon ratio, $\eta$, 
lies within the range $4.37\times10^{-10} \la \eta \la 
5.32\times10^{-10}$, implying $0.016 \la
\Omega_{\rm b}\,h^2_{100} \la 0.020$.
\end{abstract}


\section{Introduction}

Molecular hydrogen \hh, being the most abundant molecular species
in the universe, plays a central role in star formation processes
both at low and high redshifts (e.g., Shull \& Beckwith 1982).
The \hh molecule is known to be an important coolant for
gravitational collapse of hydrogen clouds at $T \sim 1000$~K.
Fractional abundances of \hh in interstellar 
\ion{H}{1} regions were calculated first by 
Gould \& Salpeter (1963) and by Hollenbach,  
Werner, \& Salpeter (1971). 
They predicted the fractional
abundance of molecular hydrogen $f_{{\rm H}_2} \simeq 10^{-3}$,
where $f_{{\rm H}_2} = 
2n_{{\rm H}_2}/(n_{\rm H} + 2n_{{\rm H}_2})$
for a standard galactic disk cloud 
with $n_{\rm H} \simeq 10$ cm$^{-3}$,
gas temperature of about 100~K,
and the optical depth to the cloud center due to dust grains
$\tau_{\rm g} \simeq 0.1$. 

The first detection of \hh in the ISM was performed by Carruthers
(1970) with a rocket-borne ultraviolet spectrometer.
Later, after the launch of the {\it Copernicus} satellite,
the far-UV absorption bands of \hh have been observed 
in the Galactic disk toward numerous relatively bright
stars with visual extinctions $A_{\rm V}$ of less than about
1 mag (e.g. Spitzer \& Jenkins 1975; Savage et al. 1977).
More recent observations of \hh with the
{\it Interstellar Medium Absorption Profile Spectrograph}
(IMAPS)
\citep{jp97}, the {\it Orbiting and Retrievable Far and
Extreme Ultraviolet Spectrometer} (ORFEUS) 
(Richter et al. 1998; de Boer et al. 1998), and the
{\it Far Ultraviolet Spectroscopic Explorer} (FUSE)
(Moos et al. 2000)
have shown the ubiquity of \hh in the disk and halo of the
Galaxy and in the Magellanic Clouds.
In the FUSE mini-survey, 
\hh was detected in the sightlines with neutral hydrogen column
densities $N$(\ion{H}{1}) $\simeq (3 \div 6)\times10^{20}$
cm$^{-2}$, the fractional abundances
$f_{{\rm H}_2} \simeq 7.4\times10^{-6} \div 3.4\times10^{-2}$ and
the excitation temperatures 
$T_{\rm ex} \simeq 180 \div 540$~K
\citep{sh2000}. 

In these observations the range of $N$(\ion{H}{1}) resembles
the range of hydrogen column densities measured in QSO
damped Ly$\alpha$ systems (DLA hereinafter) -- the systems with
$N$(\ion{H}{1})~$> 2\times10^{20}$ cm$^{-2}$, which are believed to
originate in intervening galaxies or proto-galaxies located at
cosmological distances \citep{wolfe95}.
Thus we may expect to observe far-UV \hh lines from
electronic-vibrational-rotational bands shifted 
to the blue spectral range ($\lambda > 3000$ \AA) at $z \ga 2$.
The detection of the first H$_2$-bearing cosmological cloud 
in the \zabs = 2.811 DLA system
toward Q0528--250 was made by Levshakov \& Varshalovich (1985)
and later confirmed by Foltz, Chaffee, \& Black (1988). 
Two additional \hh systems were found  
at \zabs = 1.97 toward Q0013--004 \citep{geb97} and at
\zabs = 2.34 toward Q1232+0815 \citep{geb99}. The latter was
recently confirmed by Petitjean, Srianand, \& Ledoux (2000).
The fourth identification of \hh was made at \zabs = 3.39
toward Q0000--2620 by Levshakov et al. (2000b, 2001) and a
tentative fifth \hh system was reported by Petitjean et al.
(2000) at \zabs = 2.374 toward Q0841+129.

Molecular gas in the early Universe has also been detected through
the spectral-line emission from
the tracer carbon monoxide (CO) molecule in the ultraluminous galaxy
IRAS\, 10215+4724 at $z = 2.28$ (Brown \& Van den Bout 1992;
Solomon, Downes, \& Radfort 1992), 
in the clover-leaf quasar H1413+117
at $z = 2.56$ (Barvainis et al. 1997),
in the quasar BR1202--0725 
at $z = 4.69$ (Ohta et al. 1996; Omont et al.
1996), in the quasar BRI1335--0418 at $z = 4.41$ (Guilloteau et al.
1997; Carilli, Menten, \& Yun 1999), 
in the IR luminous galaxy J14011+0252 at $z = 2.565$ (Frayer et al.
1999), in the extremely red galaxy HR10 at $z = 1.44$ (Andreani et al.
2000), in two powerful radio galaxies 4C60.07 and 6C1909+722
at $z = 3.79$ and 3.53, respectively (Papadopoulos et al. 2000), 
in the lensed quasar APM08279+5255 at $z = 3.911$ (Downes et al. 1999;
Papadopoulos et al. 2001) and in the QSO SDSS 1044--0125 at
$z = 5.8$ (Iwata et al. 2001).

However,
identification of \hh lines in the damped Ly$\alpha$ systems remains
a rather rare event. 
Most DLAs yielded only upper limits of
$f_{{\rm H}_2} \la 10^{-7} \div 10^{-6}$.
The lack of \hh in these systems might be a consequence
of the lower metal abundances, lower dust contents, 
higher temperature and
ultraviolet fluxes when compared with the ISM in the
Galaxy and in the Magellanic Clouds. Low 
molecular abundances found in high redshift DLAs resemble
the properties of, at least, one
local DLA system -- the metal-deficient ($Z/Z_\odot \sim 1/50$) 
starburst blue compact galaxy I~Zw~18 with 
$N$(\ion{H}{1}) $ \simeq 3\times10^{21}$ cm$^{-2}$ \citep{ku94}
where recent FUSE
observations set a limit of $f_{{\rm H}_2} \ll 10^{-6}$ 
\citep{vm00}.

The measured upper limits of $f_{{\rm H}_2}$ in high redshift
metal-poor DLAs are slightly lower than the 
calculated fraction of molecules in
the pristine gas -- $f_{{\rm H}_2} \simeq 10^{-6} \div 10^{-5}$
depending on cosmological models \citep{gp98}. 
Although UV photons produced by the first objects can quickly
destroy all \hh in the pristine gas and thus
delay further structure formation (Haiman, Rees, \& Loeb 1997),  
we may expect that the electrons in the
photoionized regions around first stars will in turn act
as a catalyst in the H$^{-}$ channel to produce \hh at 
redshifts $z < 50$
where the H$^{-}$ channel proceeds much more quickly than
the H$^{+}$ one \citep{fc2000}. Therefore direct measurements
of gas density, UV radiation
field, and \hh formation and destruction rates in DLA systems
may shed light on star formation processes
at very high redshift.

In this paper we present results of 
the study of \hh lines identified in the sixth DLA
system at \zabs = 3.025 toward Q0347--3819.
We also use the information from the molecular lines
together with H and metal lines to refine the
deuterium abundance determination by D'Odorico, 
Dessauges-Zavadsky, \& Molaro (2001, hereinafter DDM)  
and the values of metal abundances (DDM; Prochaska \&
Wolfe 1999).

The outline of the paper is as follows. We discuss 
observations in Section 2. 
The methods of data analysis are presented in Section 3.
The identification of \hh lines and the calculation of their
proper redshift is given in \S\,3.1. 
The broadening $b$-parameter is evaluated in \S\,3.2.
The column densities $N$(J) for the
rotational levels from J = 0 to 5 of the lowest vibrational
level v = 0 in the ground electronic state
$X\,{}^1\Sigma^+_g$
(for the \hh notation, see Field et al. 1966),
the fractional molecular abundance, and the ortho:para-\hh
ratios are calculated in \S\,3.3.
In \S\,3.4 we compute
the excitation temperature $T_{\rm ex}$. 
The results of a stacking technique applied to the measurements of
the abundances of \ion{Zn}{2}
and \ion{Cr}{2} ions and
the dust-to-gas ratio in the whole \zabs = 3.025 system are
estimated in \S\,3.5. In \S\,3.6 we analyze within one
self-consistent model
all available profiles of H, D, neutral atoms and low ions 
to infer column densities and relative abundances of D/H
and metals. In Section 4 we discuss the results and present
their further analysis. In \S\,4.1 we estimate the kinetic
temperature, $T_{\rm kin}$, and the gas volumetric density,
$n_{\rm H}$, using the
\ion{C}{2}$^\ast$/\ion{C}{2} and \ion{C}{1}/\ion{C}{2} ratios.
The distribution of \hh over the
rotational levels, the photo-dissociation rate of \hh and
the rate of the molecular hydrogen formation upon
grain surfaces are considered in \S\,4.2.
The revealed [$\alpha$-element/iron-peak] enhancement is
discussed in \S\,4.3. In Section 5 we summarize the results.

\section{Observations}

Q0347--3819 is a bright quasar ($V = 17.3$) with $z_{\rm em} = 3.23$
discovered by Osmer \& Smith (1980), which shows a damped Ly$\alpha$
system at \zabs = 3.0245. A column density of
$N$(\ion{H}{1}) = $(6.3^{+1.6}_{-1.3})\times10^{20}$ cm$^{-2}$ was
first measured by Williger et al. (1989) and later confirmed by
Pettini et al. (1994) as $N$(\ion{H}{1}) = $(5\pm1)\times10^{20}$
cm$^{-2}$. 
Q0347--3919 has a rich absorption-line spectrum 
studied by Pierre, Shaver, \& Robertson (1990),
Centuri\'on et al. (1998), Ledoux et al. (1998), and by
Prochaska \& Wolfe (1999).

High-quality data of this QSO
in the UV (3650 \AA\,$ < \lambda < 4900$ \AA)
and in the near-IR (6700 \AA\,$ < \lambda < 10000$ \AA)
ranges
were obtained during the commissioning of the UVES
on the VLT 8.2m Kueyen telescope at Paranal, Chile, in December
1999, and have been
released for public use. The UVES spectrograph is
described by D'Odorico et al. (2000) 
and details of the data reduction are given in 
Ballester et al. (2000). 
Two exposures of
4500~s and 5000~s, covering the UV and the near-IR ranges,
were obtained with a resolution FWHM $\simeq 7.0$ km~s$^{-1}$ and
$\simeq 5.7$ km~s$^{-1}$, respectively.
An average signal-to-noise ratio per pixel of
$\sim 20, 25, 40$ and 15 was achieved in the final spectrum 
at $\lambda \sim 3650$ \AA\,, 4900 \AA\,, 6700 \AA\,, and 10000 \AA\,,
respectively. 

The \zabs = 3.025 DLA exhibits a multicomponent velocity structure with
the dominating component at  \zabs = 3.024856 
as found by DDM who
discussed the first detection of the deuterium lines in DLAs.
The total \ion{H}{1} column density derived in DDM
from the best fit of Ly$\beta$ 
equals $(4.3\pm1.0)\times10^{20}$ cm$^{-2}$,
which agrees well with the measurement of Pettini et al. (1994)
from Ly$\alpha$.
The ionization structure of this DLA is also complex. There are at least
two gas components: a warm gas seen in lines of neutral atoms, H$_2$, and low ions,
and a hot gas where the resonance doublets of \ion{C}{4} and \ion{Si}{4} are
formed. These high-ion transitions, 
observed by Prochaska \& Wolfe (1999) with the Keck HIRES spectrograph show, however,
different profile shapes as compared with the low ion-transitions. The physical
properties of the 
hot gas component will not be studied in the present paper.

\section{Measurements}

All absorption lines studied in the present paper lie in the
Ly$\alpha$ forest where their analysis is complicated by
possible \ion{H}{1} contaminations and
uncertainties in a local continuum, drawn by using
`continuum windows' in the Ly$\alpha$ forest which could be located
quite far from the feature under analysis. 
This kind of uncertainties is illustrated in Figure~1 where a portion
of the Q0347--3819 spectrum with a few identified \hh lines is shown.
To achieve the highest possible accuracy 
under these circumstances, 
we have always used in parallel several different lines
to derive the physical
parameters of the DLA system.

\subsection{Redshift of the \hh system}

Wavelengths, oscillator strengths, and radiation damping constants
for the Lyman and Werner bands of the \hh molecule were taken from 
the theoretical calculations of Abgrall \& Roueff (1989). 
These data seem to be reasonably accurate as shown by
high spectral resolution IMAPS measurements of molecular
hydrogen toward the Orion belt stars 
\citep{jen2000}.

The identification of the \hh lines was started from
the subsystem which has the largest \ion{H}{1}
column density (see \S\, 3.6).
Preliminary inspection of the Q0347--3819 spectrum
revealed 88 absorption features at the expected positions of
the \hh lines. 
However, not all of them were suitable for 
further analysis due to \ion{H}{1} contaminations.
For this reason, to estimate $z_{{\rm H}_2}$ we used only those
\hh lines which show symmetric profiles with their blue and
red sides reaching the continuum. These lines are listed in
Table~1 (see also Figures~5 and 6).

Given a fit to the continuum and the line measurement
interval, the equivalent width $W_\lambda$, the line center
$\langle \lambda \rangle$ and their errors $\sigma_{W_\lambda}$ and
$\sigma_{\langle\lambda\rangle}$ can be computed from the 
following equations (e.g. Levshakov et al. 1992):
\begin{equation}
W_\lambda = \Delta\lambda\,\sum^M_{i=1}\,(1 - F_i)\; ,
\label{eq:E1}
\end{equation}
\begin{equation}
\langle \lambda \rangle = \frac{\Delta\lambda}{W_\lambda}\,
\sum^M_{i=1}\,\lambda_i (1 - F_i)\; ,
\label{eq:E2}
\end{equation}
\begin{equation}
\sigma_{W_\lambda} = \Delta\lambda\,\left[
\sum^M_{i=1}\,\sigma^2_i + \langle \frac{\sigma_c}{C} \rangle^2\,
(M - W')^2 \right]^{1/2}\; ,
\label{eq:E3}
\end{equation}
and
\begin{equation}
\sigma_{\langle\lambda\rangle} = \left[\,\frac{\Delta\lambda}{W_\lambda}\,
\sum^M_{i=1}\,(\lambda_i - \langle \lambda \rangle)^2\,
(1 - F_i) \right]^{1/2}\; ,
\label{eq:E4}
\end{equation}
where $\Delta\lambda$ is the wavelength interval between pixels, 
$M$ is the number of pixels involved in the sum, 
$F_i$ is the normalized intensity in the $i$th pixel
whose wavelength is $\lambda_i$, 
$\sigma_i$ is the noise level in the $i$th pixel,
$W' = W_\lambda/\Delta\lambda$ is the
dimensionless equivalent width in pixels, 
and $\langle \sigma_c/C \rangle$
is the mean accuracy of the continuum fit over the
line measurement interval.

The errors in $W_\lambda$ and $\langle \lambda \rangle$ are 
caused mainly by the finite number of photons counted. In addition, for
weak absorption features ($W' \ll 1$) the second term in
equation (\ref{eq:E3}) is of about the same order of magnitude
as the first one, meaning that the identification of weak \hh lines
is very sensitive to the accuracy of the local continuum fitting. 

The analyzed 35 transitions of the Lyman and Werner bands are listed
in column (2) of Table~1, whereas their J-values are given in
column (1).
The measured heliocentric position of the
center of each spectral feature and its standard deviation 
 were calculated according to
equations (\ref{eq:E2}) and (\ref{eq:E4}) and
are presented
in columns (3) and (4), respectively.
The observed equivalent widths (column 5) and their errors
(column 6) were calculated 
according to equations (\ref{eq:E1}) and (\ref{eq:E3}). 
Note that the rest frame equivalent width equals
$W_{\lambda,{\rm rest}} = W_\lambda/(1 + z_{{\rm H}_2})$.
In these calculations we set for all lines
$\langle \sigma_c/C \rangle$ = 1\%
to account for the uncertainty in the local continuum level.
A formal mean value of $\langle \sigma_c/C \rangle$
estimated from the continuum windows is less than or about 1\%. 
In particular, 
$\langle \sigma_c/C \rangle$ = 0.70\% 
for L3-0R(0) and 1.03\% for L7-0P(5) which are
shown in Figures~5 and 6, respectively.
Column (7) lists the measured redshifts for each \hh transition,
their errors $\mid \sigma_z \mid$ = 
$\mid \sigma_{\langle \lambda \rangle} \mid /\lambda_0$ 
(where $\lambda_0$ is
the rest frame wavelength from Abgrall \& Roueff 1989)
are shown in column (8). The weighted mean redshift yields
$z_{{\rm H}_2} = 3.024855\pm0.000005$, which coincides  
exactly with the redshift of the strongest main component 
in metal line profiles given by DDM.
Thus, at a given S/N ratio and spectral resolution we do not
find any systematic shifts larger than a few km~s$^{-1}$
in the radial velocities between
neutral atoms and \hh molecules in the molecular cloud at 
\zabs = 3.025. 

Previous {\it Copernicus} investigations of diffuse gas
in the Galaxy showed that the mean radial velocity of
\hh lines differs from the corresponding mean of the atomic
lines. For instance, the differences of 
$\Delta v \simeq 5$ km~s$^{-1}$ 
toward $\lambda$~Sco \citep{sj75}, 
$\Delta v \simeq 11$ km~s$^{-1}$ and 
$\simeq 6$ km~s$^{-1}$ toward, respectively, $\tau$~Sco and
$\zeta$~Oph (Spitzer, Cochran, \& Hirshfeld 1974) were detected.
At the UVES spectral resolution such differences -- in case they
were present -- 
could be easily distinguished.

\subsection{Line broadening}


The measurement of the broadening $b$\footnote{\,By definition
$b = \sqrt{2}\,\sigma_v$, where $\sigma_v$ is the 
one-dimensional Gaussian velocity dispersion of molecules
along the line of sight.} 
parameter and the calculation of the
confidence ranges for its mean value were carried out in 
two steps. 

First, a least-squares analysis
was applied to the line fitting.
We combined 17 \hh lines 
(L3-0R0, L7-0R0, W3-0Q1, L3-0P1, W0-0R2, W0-0Q2, L3-0R2,
L4-0R2, L2-0P3, L2-0R3, L4-0R3, L4-0P3, L8-0P3, L12-0R3,
L4-0P4, L8-0P4, W1-0Q5)
from Table~1 
and fitted them simultaneously 
with the individual Gaussian profiles.
These lines were chosen because they exhibit `clean',  
uncontaminated profiles.
The \hh line centers were fixed at the mean $z_{{\rm H}_2}$ 
and the theoretical profiles were
convolved with the instrumental point-spread function (PSF) which was
approximated by a Gaussian with FWHM = 7.0 km~s$^{-1}$, i.e.
$b_{\rm PSF} = 4.2$ km~s$^{-1}$.
The $\Delta \chi^2$ method (Press et al. 1992) was used to
estimate the most probable $b$ value and its error\footnote{\,In the
present paper, the estimated standard errors are slightly larger
than those calculated by the Monte Carlo method because the
pipeline resampling to constant wavelength bins introduces
a correlation between data point values,
whereas the estimated mean values of the physical parameters
remain unbiased (see Appendix A).}.
For a given current value of $b$ taken from the range 
[$b_{\rm min},b_{\rm max}$],
we minimized $\chi^2$ (normalized per degree of freedom) 
by varying the line intensities independently
(we consider $\chi^2_{\rm min}$ as a $\chi^2$ distribution
with $\nu = \cal{N} - \cal{M}$ degrees of freedom, where the
number of data points $\cal{N}$ = 192 and the number of fitted
parameters $\cal{M}$ = 6 for this particular case).
With this we tried to check possible \ion{H}{1} contaminations~:
the optimization procedure would fail to reach $\chi^2_{\rm min} 
\simeq 1$
if the chosen \hh lines were essentially
contaminated by the Ly$\alpha$ forest absorption.
The calculated $\chi^2_{\rm min}$ values
in dependence of $b$
and the $1\sigma$ and
$2\sigma$ confidence levels are shown in Figure~2. 
With $\chi^2_{\rm min} = 0.925$ the most likely value for 
$b_{{\rm H}_2}$ was found to be $2.80\pm0.45$ km~s$^{-1}$ (s.d.).

In the second step,
to check the obtained value and to infer 
the Doppler parameter independently,
we calculated an optimal curve of growth for 
the \hh lines from Table~1.
The curve is shown by the solid line in 
Figure~3.
In principle, it cannot be excluded that
some unknown absorption lines 
blend with \hh features, despite their symmetric and undisturbed
profiles. Thus, the scatter of points 
in Figure~3 can be explained by this effect as well as by errors
remaining after the continuum fitting.
It is also seen from
Figure~3 that the strongest absorption lines from J = 1 and 3  
levels are slightly saturated and therefore 
their equivalent widths
are sensitive to the assumed broadening
model. The data for these levels give the
effective Doppler width $b \simeq 3$ km~s$^{-1}$.  We also depicted
two curves for $b = 2$ km~s$^{-1}$ and 4 km~s$^{-1}$ to demonstrate
the complete consistency of $b$-values 
estimated in both approaches.

We note that Jenkins \& Peimbert (1997) have measured
an interesting phenomenon of increasing 
$b$-values for \hh lines arising from progressively higher rotational
levels in a spectrum of $\zeta$~Ori~A. 
The most probable Doppler parameters were found to be~: 
$b(0) = 0.8$ km~s$^{-1}$, 
$b(1) = 2.7$ km~s$^{-1}$, 
$b(2) = 5.3$ km~s$^{-1}$, 
$b(3) = 5.9$ km~s$^{-1}$, 
and $b(5) =  9.0$ km~s$^{-1}$.
As they noted, such profile shapes are inconsistent
with a simple model of UV optical pumping which governs the \hh
rotational populations in an optically thin, homogeneous medium. 
They suggested that these \hh profiles with changing shapes
`may represent molecules forming in the postshock gas that is
undergoing further compression as it recombines and cools'. 
A similar phenomenon may also be observed
in high redshift DLAs which 
can be associated to star forming regions where shocks are
very probable.

The \zabs = 3.025 system presents the weakest
members of the Lyman and Werner series with J = 4 and 5 
but, unfortunately, we are unable to compare their $b$-values
with those from lower rotational transitions
because our signal-to-noise ratio 
is not sufficiently high.
Thus, in the following we assume a simple one-component
model for the molecular cloud with a common broadening parameter
$b_{{\rm H}_2} = 2.80$ km~s$^{-1}$ for all molecular hydrogen lines.

\subsection{\hh abundances}

In the final fitting of each line profile we combined together the
\hh lines with the same J to yield the corresponding $N$(J)
column densities and their confidence regions. A $\chi^2$
minimization procedure was applied simultaneously 
to all lines from a given rotational level J. 
To control solutions at this stage, we added to the lines listed
in Table~1 
a few weaker transitions where we observe continuum without
any pronounced absorption features at their expected positions.
The calculated $\chi^2_{\rm min}$ values and the $1\sigma$ and
$2\sigma$ confidence levels for each individual rotational
level J are shown in Figure~4. The curves $\chi^2[N({\rm J})]$ were
computed with the fixed mean $b_{{\rm H}_2} = 2.80$ km~s$^{-1}$ and
the mean $z_{{\rm H}_2}$.
The \hh lines involved in the analysis are as follows:
L3-0R0, and L7-0R0 ($\cal{N}$ = 22); 
L3-0P1, and L14-0P1 ($\cal{N}$ = 20);
W0-0R2, W0-0Q2, L3-0R2, L4-0R2, and L5-0R2 ($\cal{N}$ = 51);
W2-0R3, L2-0R3, L2-0P3, L3-0R3, L3-0P3, L4-0R3, 
L6-0R3, L7-0P3, L8-0P3,
L12-0R3, and L15-0R3 ($\cal{N}$ = 83);
L5-0R4, L6-0R4, and L8-0P4 ($\cal{N}$ = 34); 
L7-0P5, and L12-0R5 ($\cal{N}$ = 21).
The most probable column densities and their errors at the $1\sigma$ 
confidence level are:
$N(0) = (1.90\pm0.35)\times10^{13}$ cm$^{-2}$,
$N(1) = (1.60\pm0.18)\times10^{14}$ cm$^{-2}$,
$N(2) = (5.40\pm0.35)\times10^{13}$ cm$^{-2}$,
$N(3) = (1.22\pm0.05)\times10^{14}$ cm$^{-2}$,
$N(4) = (2.60\pm0.43)\times10^{13}$ cm$^{-2}$, and
$N(5) = (2.90\pm0.70)\times10^{13}$ cm$^{-2}$.
From that we calculate the total column density in \hh,
$N({\rm H}_2) = \sum_{\rm J}\,N({\rm J})$ =
$(4.10\pm0.21)\times10^{14}$ cm$^{-2}$,
and the ratio of the
total column densities in the ortho- and para-\hh states
$N_{\rm ortho}/N_{\rm para} = 3.1\pm0.3$ (s.d.). 
This result can be compared with two other molecular clouds at
\zabs = 2.8112 (Q0528--250, Srianand \& Petitjean 1998) 
and \zabs = 2.33771 (Q1232+0815, Srianand et al. 2001)
where $N$(J) values were also measured for different J.
Using the published $N$(J) values and their errors, we found
$N_{\rm ortho}/N_{\rm para} = 3.1\pm0.8$ (s.d.) at \zabs = 2.8112
and
$N_{\rm ortho}/N_{\rm para} = 2.4\pm0.9$ (s.d.) at \zabs = 2.33771.

According to our calculations presented in Section 3.6, 
the total \ion{H}{1} column density in the \zabs = 3.025 system is
$(4.225\pm0.045)\times10^{20}$ cm$^{-2}$. 
With 
$N({\rm H}_2) = (4.10\pm0.21)\times10^{14}$ cm$^{-2}$, the
ratio of H nuclei in molecules to the total H nuclei is
$$
f_{{\rm H}_2} = \frac{2\,N({\rm H}_2)}{N({\rm H})} =
(1.94\pm0.10)\times10^{-6}\; .
$$

To test the consistency of the estimated physical parameters
with the observed absorption features, we
calculated theoretical profiles for all 88 \hh lines
using the derived mean $z_{{\rm H}_2}$,
$b_{{\rm H}_2}$, and $N$(J) values. Besides we calculated 8
profiles for \hh transitions where no absorption
features have been observed
at the continuum level. The latter are 
L3-0R4, L9-0R4, L4-0R5, L9-0R5, L11-0P5, 
W0-0Q5, L6-0P6, and L7-0R7. The last two of them
give upper limits to $N(6) < 2.0\times10^{13}$ cm$^{-2}$ 
($1\sigma$ c.l.), and to
$N(7) < 8.0\times10^{12}$ cm$^{-2}$ ($1\sigma$ c.l.).

The observational and over-plotted theoretical \hh profiles
are shown in Figures~5 and 6. In these figures possible blends 
with other lines from the \zabs = 3.025 DLA and from the metal systems
listed in Appendix~B are also indicated. 
We found only one large inconsistency between 
observable and theoretical profiles,
namely for the W2-0Q(3) line (Figure~6). 
But in this case a very wide absorption trough between 3901 \AA\, and
3921 \AA\, redward the redshifted position of this molecular
line ($\lambda_{\rm obs} = 3900.3$ \AA) prevents the 
accurate estimation of the local continuum level. 
Uncertainties in the local continuum level calculations
may also cause over-absorption in the synthetic profiles as
compared with the observable
profiles seen in W1-0R0+R1, W2-0Q1 (Figure~5), W0-0Q3, L3-0P3,
L4-0P3, L10-0R3, L12-0R3, W0-0Q5, and L4-0R5 (Figure~6).  
In all other cases
the synthetic \hh profiles are in agreement with
the absorption-line features identified with the
\hh lines in the UV spectrum of Q0347--3819.  

We have also tried to detect
the \hh molecules in the blue component 
at $\Delta v \simeq -17$ km~s$^{-1}$ (see \S\, 3.6) 
with respect to the main H$_2$-bearing component 
but we found only an upper
limit. The applied method is similar to that employed by Levshakov
et al. (1989) to improve the detectability of weak carbon monoxide
(CO) UV lines in damped Ly$\alpha$ systems. 

The procedure consists of shifting the observed spectral regions
containing expected positions of the \hh transitions, which are
not contaminated by the apparent Ly$\alpha$ forest absorption, 
into the relative velocity scale centered at
$z = 3.024637$, -- the redshift of the blue subcomponent of the
dominant metal absorption in the \zabs = 3.025 DLA,
and of the consequent stacking of these regions.
Before co-adding each section, the data are rebinned 
with a bin size of 1.45 km~s$^{-1}$.
The spectra were then averaged with weights in accord with
the squares of the expected relative line strengths divided
by the respective signal-to-noise ratios. 

The result obtained from this stacking procedure is shown in Figure~7.
The composite spectrum formed by the addition of 18 \hh bands
(with no pronounced absorption features in the range
$\Delta v = \pm6$ km~s$^{-1}$)
which are labeled in the corresponding panels is depicted in the
bottom panel. The resulting co-added data have an average 
S/N $\simeq 100$
and show no absorption at $ v = 0$ km~s$^{-1}$, whereas the blue
wing of the composed molecular line from the $z = 3.024855$ system
is clearly seen at $v \simeq +16$ km~s$^{-1}$.

The composite spectrum gives $W_{\lambda,{\rm rest}} < 0.6$ m\AA\,
$(3\sigma\, {\rm c.l.})$ for the equivalent width of the sum of the \hh
bands in the rest frame of the absorber. This limit is sensitive,
in general, to errors in continuum fitting. In our case
the resulting profile between $-6$ km~s$^{-1}$ $< v <$ 6 km~s$^{-1}$
has no apparent deviations from unity, the estimated mean value
is $\langle C \rangle = 1.003\pm0.003$ and thus 
$\langle \sigma_c/C \rangle  = 0.003$. Substituting these values
in equation (3) we obtain the limiting equivalent width ($W' \ll 1$)
$\sigma_{\rm lim} = 0.2$ m\AA\, and $W_{\lambda,{\rm rest}} =
0.6$ m\AA\ ($3\sigma_{\rm lim}$). 

For a weak absorption line with the central optical depth
$\tau_0 \ll 1$, the column density is independent of the
broadening parameter $b$ and is linearly proportional to the
equivalent width $W_\lambda$:
\begin{equation}
N = 1.13\times10^{20}\,\frac{W_\lambda}{\lambda^2_0 f},\, 
{\rm cm}^{-2}
\label{eq:E7}
\end{equation}
where $f$ is the oscillator strength for the absorption
transition, $\lambda_0$ is the line center rest wavelength in \AA,
and $W_\lambda$ is the rest equivalent width in \AA.

Using this linear approximation to the curve of growth, we
calculate the composite equivalent width $W_{\rm J}$
for the \hh bands from the same J level
\begin{equation}
W_{\rm J} = a\,N_{\rm J}\,\sum_i\,\lambda^2_{{\rm J},i} f_{{\rm J},i}\; ,
\label{eq:E8}
\end{equation}
where $a = 8.85\times10^{-21}$.

\noindent
Then the total equivalent width for all \hh bands presented in 
Figure~7 is
\begin{equation}
W_{\Sigma} = \sum_{\rm J}\,W_{\rm J} = a\,\sum_{\rm J}\,
\epsilon_{\rm J}\,N_{\rm J} = a\,\hat{\epsilon}N_{{\rm H}_2}\; ,
\label{eq:E9}
\end{equation}
where $\epsilon_{\rm J} = \sum_i\,\lambda^2_{{\rm J},i} f_{{\rm J},i}$,\,
$\hat{\epsilon}$ is an effective value (in our estimation
$\hat{\epsilon} = {\rm max}\,\epsilon_{\rm J}$), and
$N_{{\rm H}_2} = \sum_{\rm J}\,N_{\rm J}$.

\noindent
The coefficients $\epsilon_{\rm J}$ do not vary significantly and
for the \hh bands from Figure~7 
$\epsilon_0 = 6.1\times10^4$,
$\epsilon_1 = 5.5\times10^4$,
$\epsilon_2 = 5.3\times10^4$, and
$\epsilon_3 = 9.8\times10^4$. Thus, the aforementioned upper limit
$W_\lambda < 0.6$ m\AA\, and equation (\ref{eq:E9}) yield
$N({\rm H}_2) < 7.0\times10^{11}$ cm$^{-2}$ ($3\sigma$ c.l.) and
$f_{{\rm H}_2} < 3.3\times10^{-9}$ ($3\sigma$ c.l.), 
which is well below that found in Galactic \ion{H}{1} clouds. 
It means that molecular hydrogen is not homogeneously distributed
in the \zabs = 3.025 absorbing region.

The measured $N_{\rm ortho}/N_{\rm para}$ ratios 
in the DLAs at \zabs = 3.025 (Q0347--3819), 
2.8112 (Q0528--250), and 2.33771 (Q1232+0815)
are equal to 3 (within the uncertainty interval), which is 
just the equilibrium value for the ortho:para ratio at the hot
formation of \hh on grain surfaces in relatively transparent
clouds (e.g., Spitzer \& Morton 1976). 
In the pristine gas, the ortho:para \hh ratio decreases
from the equilibrium value of 3 at $z \simeq 100$ to its 
`freeze-out' quantity of 0.25 at $z < 20$ (Flower \&
Pineau des For\^ets 2000).
Thus, we can conclude that \hh observed in these DLAs
has undergone a re-forming phase
following photo-dissociation of the pristine \hh.

\subsection{Excitation temperature}

In the ground electronic-vibrational state, all the rotational
levels with odd J are nuclear triplet states ( ortho-\hh), and
all those with the even J 
are singlet states ( para-\hh). Thus, for the ortho-\hh
the statistical weight, $g$(J), of a level J 
equals 3(2J+1), whereas for the para-molecules 
$g({\rm J}) = (2{\rm J} + 1)$.
In thermal equilibrium, 
the collisionally dominated
distribution of 
molecules over different J levels is given by a Boltzmann 
law with an effective excitation temperature $T_{\rm ex}$
\begin{equation}
\frac{N({\rm J})}{N(0)} = \frac{g({\rm J})}{g(0)}\,
\exp\,\left[ - \frac{B_{\rm v}{\rm J(J+1)}}{T_{\rm ex}}\right]\; ,
\label{eq:E6}
\end{equation}
where the constant $B_{\rm v}$ equals 85.36~K for the vibrational
ground state.

This equation shows that $T_{\rm ex}$ is proportional to the
negative inverse of the slope of a line 
(called `the excitation diagram')
drawn through the
points of the respective J levels in 
a plot log~[$N$(J)/$g$(J)] versus $E$(J), where 
$N$(J) is the column density measured in the rotational level J,
$E$(J) is the excitation energy of this level relative to J = 0,
and $g$(J) is its statistical weight. 
Figure~8 shows this diagram for different J levels. A least-squares
fit, weighted by $1/\sigma^2$ for each log~[$N$(J)/$g$(J)],
yields $T_{\rm ex} = 825\pm110$~K (s.d.) and 
log~$N(0) = 13.31\pm0.03$ in cm$^{-2}$\, (s.d.). 
This value of $N(0)$ agrees well with what has been measured from
profile fitting in \S\, 3.3.
Thus, we find that the population of the six lowest J levels
can be fitted with a single Boltzmann distribution.

The unique $T_{\rm ex}$ for all levels from J = 0 to 5 may
imply that the \hh rotational population is dominated
by collisions in a gas whose kinetic temperature is about 800~K.
The kinetic temperature in the H$_2$ cloud can be estimated from
the widths of the \hh and metal lines.
In \S\,3.6, we show 
that profiles of all metals from \ion{C}{2} to \ion{Zn}{2}  
can be described with the same broadening parameter of 
$b_{\rm Me} = 2.71\pm0.05$ km~s$^{-1}$.  
With $b_{{\rm H}_2} = 2.80\pm0.45$ km~s$^{-1}$, it gives an
upper limit of $T_{\rm kin} \leq 430$~K 
in the main H$_2$-bearing component.
The fact that $T_{\rm kin}$ is less than  $T_{\rm ex}$ indicates that the 
population of the low J levels is caused not only by collisions but 
by UV-pumping as well.
The J = 0 and 1 rotational levels,
being usually populated in dense molecular clouds 
by thermal proton collisions, 
may not be in equilibrium with the kinetic temperature in
lower density clouds which are not optically thick in 
the L0-0R(0), R(1), P(1) and W0-0R(0), R(1), Q(1) lines. 
In low density clouds
the self-shielding effects are negligible and do not
reduce significantly UV-pumping in these lines
(e.g., Shull \& Beckwith 1982). 
For the further analysis we set $T_{\rm kin} = 400$~K to be consistent
with the measured $b$-values.

\subsection{Cr/Zn and dust content} 

{\it Copernicus} \hh survey has shown that 
in our Galaxy the
molecular fraction $f_{{\rm H}_2}$ is correlated with $E(B-V)$ and
with the total neutral hydrogen column density
(e.g. Spitzer \& Jenkins 1975).
The presence of dust grains in damped Ly$\alpha$ systems is usually
estimated from the abundance ratio [Cr/Zn]\footnote{\,[X/Y] = (X/Y) --
(X/Y)$_\odot$, where (X/Y) is the logarithmic value of the element
ratio by number without reference to the solar value. Solar abundances
(X/Y)$_\odot$ are taken from Grevesse \& Sauval (1998).}
assuming that Zn is undepleted (Pettini et al. 1994).

Preliminary visual inspection of the red part of the Q0347--3819
spectrum has shown weak absorption features at the expected
positions of \ion{Zn}{2} and \ion{Cr}{2} lines. In order to increase
sensitivity and to measure the abundance ratio [Cr/Zn] we applied
a stacking procedure similar to the one described in \S\,3.3 
to the \ion{Zn}{2} $\lambda 2026.136$ \AA\,
and $\lambda 2062.664$ \AA\, lines, and to the \ion{Cr}{2}
$\lambda 2056.254$, \AA\, $\lambda 2062.234$, \AA\, and 
$\lambda 2066.161$ \AA\, lines.
The wavelengths and the oscillator strengths for these lines are
taken from Morton (1991), and from 
Bergeson \& Lawler (1993), respectively. 
The resulting composite spectra of \ion{Zn}{2} and \ion{Cr}{2}
were included in the metal analysis in the next \S\,3.6
(see Figure~10 and Tables~3 and 4).
In this Figure, 
two pronounced absorption features at
$\Delta v \simeq +6$ km~s$^{-1}$ (\ion{Cr}{2}) and at
$\Delta v \simeq +13$ km~s$^{-1}$ (\ion{Zn}{2}) are telluric lines.
The former partly blends 
the red wing of the \ion{Cr}{2} line, therefore
we do not use this portion of the \ion{Cr}{2} profile in the fitting
procedure. 
The estimated total column densities are
$N$(\ion{Zn}{2}) = $(6.19\pm0.51)\times10^{11}$ cm$^{-2}$,
and
$N$(\ion{Cr}{2}) = $(2.76\pm0.22)\times10^{12}$ cm$^{-2}$.
With these numbers one finds
[Cr/Zn] = $-0.37\pm0.06$.

In Table~2 we summarize the results of the abundance measurements
of molecular hydrogen, [Cr/Zn] (and/or [Fe/Zn]), and neutral
hydrogen in the DLAs available up to now. The presence of dust
in these DLAs and the relation between the fractional molecular
abundances and relative heavy element depletion is illustrated
in Figure~9. The measured data for the five extragalactic 
molecular clouds (marked by numbers in Figure~9) can be fitted
with a linear law~: $\log f_{{\rm H}_2} = A\,{\rm [Cr/Zn]} + B$,
with $A = -6.053\pm0.116$ and $B = -7.888\pm0.051$. 
The linear regression
is shown in Figure~9 by the solid line\footnote{\,For the published  
27 DLAs where all three elements \ion{Zn}{2}, \ion{Cr}{2}, and
\ion{Fe}{2} were measured,   
[Cr/Fe] = $0.12\pm0.13$ (s.d.). 
This implies that Fe is slightly more 
depleted than Cr assuming both elements track 
each other in the abundances and,
thus, the slope in our regression analysis may be slightly steeper.
However, the accuracy of the 
[Cr/Zn] and [Fe/Zn] measurements is not sufficiently high to provide
an unambiguous renormalization of  
one element to the other. Therefore 
[Fe/Zn] and [Cr/Zn] are used interchangeably in the present calculations.}.
Again, similarly to 
our previously obtained results (Levshakov et al. 2000b), we find
a strong correlation between $\log f_{{\rm H}_2}$ 
in the DLA systems with detectable \hh absorption
and the heavy element depletion. 
The presence of upper limits below the linear regression line
may be explained 
by the suggestion that in these systems a relatively high 
temperature of dust grains prevents molecular formation.

The degree of the heavy element depletion may be used to
estimate the dust-to-gas ratio, $\tilde{k}$, if two measured
elements have the same nucleosynthesis history (cf. Vladilo 1998)~:
\begin{equation}
\tilde{k} = \frac{10^{{\rm [Zn/H]}_{\rm obs}}}
{f_{{\rm Zn, ISM}} - f_{{\rm Cr, ISM}}}\,
\left( 10^{{\rm [Cr/Zn]}_{\rm obs}} - 1\right)\; ,
\label{eq:E10}
\end{equation}
where the fraction in dust $f_{{\rm Zn, ISM}} = 0.35$ and 
$f_{{\rm Cr, ISM}} = 0.92$
refers to the Galactic interstellar medium.

Error propagation method applied to the column density
measurements yields [Zn/H]~= $-1.503\pm0.054$, 
and then Eq.~(\ref{eq:E10})
results in the average value for the whole DLA
$\tilde{k}_{\rm DLA} = 0.032\pm0.005$ ($1\sigma$ c.l.)
showing that
the dust content in the \zabs = 3.025 absorber is approximately
30 times lower as compared with the Milky Way.
If, however, we consider the same quantity for the H$_2$-bearing cloud
using the data from Table~3, then  $\tilde{k} \simeq 0.047$, i.e.
the dust content in the molecular cloud is about 1.5 times higher 
as compared with the average value.

\subsection{D/H and metal abundances}

In this section we study metal profiles and
derive a new D/H ratio by fitting simultaneously
the available Lyman series lines and metals.
The knowledge of the proper distribution of the velocity
components along the line of sight is especially crucial
for the D/H measurements, since
\ion{D}{1} lines are always blended with the
blue wings of the corresponding \ion{H}{1} lines
(for numerous examples, see Levshakov, Kegel, \& Takahara 1999 and
references cited therein).

In the UVES spectrum of Q~0347--3819, 
at \zabs = 3.025 we have identified 
absorption lines of
\ion{C}{2}, \ion{N}{1}, \ion{O}{1}, \ion{Si}{2}, \ion{P}{2}, 
\ion{Ar}{1}, \ion{Cr}{2}, \ion{Fe}{2}, \ion{Zn}{2}, 
and \ion{C}{2}$^\ast$.
The identified metal absorption lines are shown in Figure~10,
and in Figure~11 -- the atomic hydrogen lines. 
The atomic data were taken mainly from 
Morton (1991) with a few additions of new results for 
\ion{Si}{2}\, from Spitzer \& Fitzpatrick (1993) and
Charro \& Mart\'in (2000), for \ion{Ar}{1}\, from
Federman et al. (1992), for \ion{Cr}{2}\, and \ion{Zn}{2}\,
from Bergeson \& Lawler (1993), for \ion{Fe}{2}\, from
Cardelli \& Savage (1995), Raassen \& Uylings (1998), and
Howk et al. (2000).

To construct a model for the radial velocity distribution, 
we firstly analyzed the profiles of each
individual atom/ion to select lines (or parts of them) which are
less contaminated and disturbed by the Ly$\alpha$ forest absorption.
The chosen profiles and/or  their components are marked in Figures~10
and 11 by horizontal lines at the bottoms of the panels
(these horizontal lines also show the pixels involved in the 
final optimization procedure). Other spectra in
Figures~10 and 11, not marked by the horizontal lines, 
were used to control consistency of the obtained results.

Our model is based on the following assumptions: ($i$) the whole DLA is
considered as a continuous absorbing region with fluctuating gas density
and velocity fields. ($ii$) The metallicity for each element is set constant
through the DLA. ($iii$) Some of the observed species may have similar
profiles, i.e. their fractional ionization ratios remain constant along
the sightline.
In this case complex profiles can be modeled using the sum of the same
number of components with identical centers and equal relative intensities
for any pair of lines (Levshakov et al. 1999b).

The line profiles are described
by the sum of $m$ Voigt functions~:
\begin{equation}
\tau_\lambda = \sum^m_{i=1}\,N_i\,{\cal V}
[(\lambda - \lambda_0 - \lambda_0 \frac{v_i}{c})/
\Delta\lambda_i ]\; ,
\label{eq:E11}
\end{equation}
where $\tau_\lambda$ is the optical depth at wavelength $\lambda$
within the line profile, $N_i$ is the column density in the $i$th
component, $v_i$ is its Doppler shift, and 
$\Delta\lambda_i = \lambda_0\,b_i/c$ 
is its Doppler width with $c$
being the speed of light and $\lambda_0$ the line center.

Equation (\ref{eq:E11}) requires additional suggestion about the
broadening mechanism. We assume that the main broadening is caused
by bulk motion. For the component seen in the \hh lines this
suggestion is supported by the measured equivalence of
$b_{{\rm H}_2}$ and $b_{\rm Me}$,
whereas for other components it is a model assumption which has to be
checked during the optimization procedure.

Now we can describe our model in detail. It is fully defined
by specifying $\{ N_i \}^\kappa_{i=1}$ -- the column density
of the 1st components for each ion, 
with $\kappa$ being the total number of ions~;
$\{ {\cal R}_{1i} \}^m_{i=2}$ -- the relative intensities of the
subcomponents, where for a given ion
${\cal R}_{1i} = N_i/N_1$~;
$\{ v_i \}^m_{i=1}$ -- the Doppler shifts~; and
$\{ b_i \}^m_{i=1}$ -- the Doppler widths.
All these parameters are components of the parameter vector $\theta =
\{\theta_1, \theta_2, \ldots, \theta_q\}$.
In addition to $\theta$, the D/H ratio is to be set. To estimate
$\theta$ from the absorption line profiles, we minimize the
objective function
\begin{equation}
\chi^2({\rm D}/{\rm H}) = \frac{1}{\nu}\,\sum^L_{\ell=1}\,
\sum^{M_\ell}_{i=1}\,
\left[ {\cal F}^{\rm obs}_{\ell, \lambda_i} - 
{\cal F}^{\rm cal}_{\ell, \lambda_i} (\theta) \right]^2 /
\sigma^2_{\ell, i}\; ,
\label{eq:E12}
\end{equation}
where D/H is the total hydrogen isotopic ratio for the
whole system.
In (\ref{eq:E12}), 
${\cal F}^{\rm obs}_{\ell, \lambda_i}$ is the observed 
normalized intensity of the spectral line $\ell$, and $\sigma_{\ell, i}$
is the experimental error within the $i$th pixel of the line
profile. 
${\cal F}^{\rm cal}_{\ell, \lambda_i}(\theta)$ is 
the calculated intensity of line $\ell$ at the same $i$th pixel
having wavelength $\lambda_i$
\begin{equation}
{\cal F}^{\rm cal}_{\ell, \lambda_i} (\theta) = \int\,
{\rm e}^{-\tau_{\ell, \lambda_i}(\theta)}\,\phi_{\rm PSF}
(\lambda_i - \lambda)\,d\lambda\; ,
\label{eq:E13}
\end{equation}
where $\tau_{\ell, \lambda_i}(\theta)$ is defined through
Eq.(\ref{eq:E11}). The instrumental point-spread function
$\phi_{\rm PSF}(\lambda)$ is approximated by a Gaussian with
FWHM = 7.0 km~s$^{-1}$ (see \S\, 3.2).

\noindent
The total number of spectral lines involved in the optimization
procedure is labeled by $L$, and the total number of data
points ${\cal M} = \sum^L_\ell M_\ell$, where $M_\ell$ is the number
of data points for the $\ell$th line. The number of degrees
of freedom is labeled by $\nu$.

After preliminary trial computations we were able to separate all
available lines into two groups consisting of similar profiles
(the fitting of each group of lines yielded acceptable
$\chi^2$ values). 
Similarity in this context means that
the density weighted radial velocity distribution functions
for elements combined in a group are equal 
[see eq.(\ref{eq:E14a}) and explanations below].
First group includes $\kappa = 9$ different ions
(\ion{H}{1}, \ion{C}{2}, \ion{C}{2}$^\ast$, \ion{N}{1},
\ion{O}{1}, \ion{Si}{2}, \ion{P}{2}, \ion{Ar}{1}, and \ion{Zn}{2})
with the total number of data points ${\cal M}$ = 566 
and $\nu$ = 544 (we consider
\ion{C}{2}$^\ast$ as a separate element since it has 
a column density different from \ion{C}{2}).
The second one consists of
$\kappa = 2$ ions (\ion{Fe}{2} and \ion{Cr}{2}) 
with  ${\cal M}$ = 110, $\nu$ = 99.
These two groups are essentially the refractory and
non- (or little) refractory elements.
 
We do not consider the absorption features seen at
$\Delta v \simeq 40$ and 60 km~s$^{-1}$ in the
\ion{C}{2} $\lambda 1036.3$ \AA\,,
\ion{N}{1} $\lambda 1134.2$ \AA\,,
\ion{O}{1} $\lambda 1039.2$ \AA\,, and 
\ion{Si}{2} $\lambda 1193.3$ \AA\, profiles (Figure~10)
since we are mostly interested in the main H$_2$-bearing component 
($\Delta v = 0$ km~s$^{-1}$) and in the blue side parts
of the profiles which restrict the configuration of the
velocity field important for the analysis of the observed  
\ion{D}{1} lines.
For this reason only blue parts of the hydrogen \ion{H}{1} Ly$\alpha$, 
Ly-5, Ly-8, Ly-10, and Ly-12 lines were included in
the analysis (see Figures~11 and 12).
The components at $\Delta v \simeq 40$ and 60 km~s$^{-1}$
increase the total column densities measured in the range
$-50$ km~s$^{-1}$ $\leq v \leq 80$ km~s$^{-1}$ at about
8\% only.

The estimation of the parameter vector $\theta$ by means of
the $\chi^2$-minimization is
a typical {\it ill-posed} problem (Tikhonov \& Arsenin 1977)
because of the presence of two
smoothing operators in (\ref{eq:E11}) and (\ref{eq:E13})
which prevent from deducing a unique value of $\theta$.
To stabilize the solution, we augment $\chi^2$ by a regularization term
in a way similar to that discussed 
in Levshakov et al. (1999b) --
instead of minimizing $\chi^2$, defined in (\ref{eq:E12}), we now seek
the minimum of another objective function, 
\begin{equation}
{\cal L}_\alpha(\theta) =
\chi^2(\theta) + \alpha\,\psi(\theta)\; ,
\label{eq:E14}
\end{equation}
where $\psi$ is a penalty function and $\alpha$ the regularization
parameter. 

The choice of $\psi$ for nonlinear problems occurs rather heuristically.
For our particular DLA
we used additional information
obtained from the analysis of \hh. Namely,
we searched for the minimum of ${\cal L}_\alpha(\theta)$ in the vicinity of
the main component (i.e. $\Delta v_{{\rm H}_2} \simeq 0$ km~s$^{-1}$) 
which is supposed to have $b \simeq b_{{\rm H}_2}$.
Then the easiest choice for $\psi$ will be
\begin{equation}
\psi = \left( \frac{b - b_{{\rm H}_2}}{\sigma_{b_{{\rm H}_2}}} \right)^2 +
\left( \Delta v - \Delta v_{{\rm H}_2} \right)^2\; ,
\label{eq:E15}
\end{equation}
and we set the regularization parameter $\alpha = 1$.

The importance of the regularization 
is illustrated in Figure~13, where 
different velocity distribution functions $p(v)$ are shown.
The differences in shapes of $p(v)$ may cause different D/H values.
For instance, being analyzed separately, the H+D profiles
yield a shallow and smoother $p(v)$ (the dashed curve in Figure~13),
whereas a simultaneous fitting with metals reveal a sharp spike
at $\Delta v \simeq 0$ km~s$^{-1}$ 
(the solid curve) and a slightly different   
structure of the blue wing of $p(v)$ at $\Delta v < -5$ km~s$^{-1}$.
The difference in D/H is, indeed, essential: D/H = 
$(2.24\pm0.67)\times10^{-5}$ in the former case (see DDM), and D/H =
$(3.75\pm0.25)\times10^{-5}$ in the latter one (see Figure~14).

The estimated column densities together with their errors 
and relative abundances in the H$_2$-bearing cloud  
are listed in Table~3. 
The errors were calculated by using the $\Delta \chi^2$ method.
They reflect internal uncertainties of the physical
parameters within the adopted model. 
Total column densities (calculated in the range
$-50$ km~s$^{-1}$ $\leq v \leq 20$ km~s$^{-1}$)
for
all elements, their relative abundances and corresponding errors
are listed in Table~4.
Besides, we also estimated 
in the main component $1\sigma$ upper limits to
$N$(\ion{C}{1}) $< 4.2\times10^{11}$ cm$^{-2}$ from
the \ion{C}{1}\,$\lambda 1277.2$ \AA\, line, to
$N$(\ion{O}{1}$^\ast$) $< 8.2\times10^{11}$ cm$^{-2}$ from
the \ion{O}{1}$^\ast$\,$\lambda 1304.8$ \AA\, line,
and to
$N$(\ion{Si}{2}$^\ast$) $< 2.6\times10^{12}$ cm$^{-2}$ from
the \ion{Si}{2}$^\ast$\,$\lambda 1533.4$ \AA\, line
using the Keck telescope spectrum of 
Q0347--3819 obtained by Prochaska \& Wolfe (1999).
These data are included in Table~3.
The most probable $b$ parameters, corresponding to the  
column densities from Table~3,
are as follows: $b_{\rm H\,I + Me} = 2.71\pm0.05$ km~s$^{-1}$, and
$b_{\rm Cr\,II + Fe\,II} = 2.80\pm0.26$ km~s$^{-1}$.
It shows that in spite of the difference in shapes of $p(v)$
for two groups of ions,
the broadening parameter in the central H$_2$-bearing component
is the same.

The measured value of log\,$N$(\ion{H}{1}) =
$20.626\pm0.005$ cm$^{-2}$ is in good agreement with the DDM
measurement of  log\,$N$(\ion{H}{1}) =
$20.63\pm0.09$ cm$^{-2}$.
The re-calculated value of D/H = $(3.75\pm0.25)\times10^{-5}$ 
is consistent with standard big bang 
nucleosynthesis predictions (e.g. Burles et al. 1999) 
if the baryon-to-photon ratio, $\eta$, is
in the range $4.37\times10^{-10} \la \eta \la 
5.32\times10^{-10}$. 
With the present-day photon density
determined from the CMBR (Mather et al. 1999), 
we can estimate the 
present-day baryon density
$\Omega_{\rm b}\,h^2_{100} \equiv 
3.73\times10^7\,(T_{\rm CMBR}/2.75)^3\,\eta\, 
\simeq 0.016 \div 0.020$.

Our new D/H ratio is shifted toward the higher 
hydrogen isotopic ratio found in the Lyman Limit Systems (LLS)
by Tytler et al. (2000),
D/H = $(3.5\pm0.5)\times10^{-5}$.
We recall that the ratio of D/H $\simeq 4\times10^{-5}$ was 
found from the Tytler et al. LLSs 
using a different mesoturbulent method by Levshakov, Kegel, \& Takahara (1998)
and by Levshakov, Tytler, \& Burles (2000). 
We thus summarize that when all most accurate measurements of D/H in high
redshift QSO absorbers are considered, one finds the same D/H value
which is about 2.5 times the mean ISM value of 
D/H = $(1.50\pm0.10)\times10^{-5}$ (e.g. Linsky 1998).

To conclude this section, we comment on the fitting of the \ion{C}{2}
$\lambda1036$ \AA\, line: the exact measurement of the column density for 
this ion is very important since the calculation of the 
volumetric gas density $n_{\rm H}$ in the \zabs = 3.025 molecular cloud is
based on the ratio of $N$(\ion{C}{2}$^\ast$)/$N$(\ion{C}{2}).
The \ion{C}{2} $\lambda1036$ \AA\ line is saturated and having 
this line alone we would be able to estimate only a lower boundary for 
$N$(\ion{C}{2}). 
But the system under study exhibits a plenty of other metal profiles
which are unsaturated.
Being fitted together with \ion{C}{2}, these profiles allow us to restrict 
the shape of the radial velocity distribution $p(v)$ and then to 
determine the real column density of \ion{C}{2}. 
As already stated above, the measurement of $N$(\ion{C}{2}) from the 
saturated $\lambda1036$ \AA\, line  is possible 
if the density weighted velocity distribution functions of \ion{C}{2} and
a match ion (which shows unsaturated absorption profiles) are equal.

Indeed, for a fixed value of $\lambda$ (or $v$ in the velocity space)
within the line profile, the optical depth $\tau_{{\rm a},i}(\lambda)$ of element
`a' in the $i$th ionization stage is given by [cf. eqs. (24)-(26) in Levshakov,
Agafonova, \& Kegel 2000]~:
\begin{eqnarray}
\tau_{{\rm a},i}(\lambda) =   
k_0\,N_{{\rm a},i}\,\int^1_0\,
n_{\rm H}(x)\,\frac{\Upsilon_{{\rm a},i}[U(x)]}{\bar{\Upsilon}_{{\rm a},i}}\,\times 
& &  \nonumber \\
\Phi_\lambda[\Delta\lambda_{\rm D}(x),v(x)]\,dx\; , & &
\label{eq:E14a}
\end{eqnarray}
where $k_0$ is a constant for a particular line, $N_{{\rm a},i}$ the column density,
$n_{\rm H}(x)$ the volumetric gas density at point $x$,
$\Upsilon_{{\rm a},i} = n_{{\rm a},i}/n_{\rm a}$ the fractional ionization,
$U = n_{\rm ph}/n_{\rm H}$ the ionization parameter with $n_{\rm ph}$ being
the number of photons with energies above one Rydberg,
$\Phi_\lambda$ the profile function at the local velocity $v(x)$,
and $\bar{\Upsilon}_{{\rm a},i}$ the mean density-weighted fractional ionization.

\noindent
If the line broadening is caused by bulk motions and the ratio of the
fractional ionizations for a pair of ions `a,$i$', `b,$j$' remains the same for
all $U(x)$ along the sightline, then
\begin{equation}
\frac{\tau_{{\rm a},i}(\lambda)}{\tau_{{\rm b},j}(\lambda)} \propto
\frac{N_{{\rm a},i}}{N_{{\rm b},j}} = {\rm constant}\; .
\label{eq:E14b}
\end{equation}
Given the integral part of (\ref{eq:E14a}), which can be estimated from the unsaturated 
lines, we can easily calculate
the column density for even saturated lines.

To test  how accurately \ion{C}{2} traces the distributions of other elements, 
we have studied the ionization properties of the warm \ion{H}{1} gas assuming
that it is embedded in an ionizing radiation field of given spectrum and
intensity.

The CLOUDY code (Ferland 1996) was used to compute the fractional ionization
of ions $\Upsilon_{{\rm a},i}$ as a function of
the ionization parameter $U$.
For a given value of $n_{\rm ph}$, 
the fluctuation of $U$ along the line of sight represents
the reciprocal hydrogen density.
We consider two limiting types of radiation fields: a hard, QSO-dominated
spectrum representative of the radiation field external to the 
H$_2$-bearing cloud at $z = 3$ (Haardt \& Madau 1996), and a soft, 
starlight background given by Mathis, Mezger, \& Panagia (1983).
For the chosen ionizing spectra, $n_{\rm ph} = 2.2\times10^{-5}$ cm$^{-3}$
and $2.0\times10^{-6}$ cm$^{-3}$, respectively. Calculations were performed
for metallicity $Z = 0.1Z_\odot$.

We found that most concordant with
\ion{C}{2} is the ion \ion{Si}{2} which traces 
\ion{C}{2} over a rather wide range of $U$.
This is illustrated in Figure~15 where the ratios 
$\Upsilon_{\rm Si\,II} / \Upsilon_{\rm C\,II}$ and  
$\Upsilon_{\rm Fe\,II} / \Upsilon_{\rm C\,II}$ (to stress the difference)
are shown as functions of $U$.  
The displayed boundaries of $U$ 
correspond to $n_{\rm H} = 0.22 \div 2.2$ cm$^{-3}$ (hard ionizing spectrum)
and to 
$n_{\rm H} = 0.02 \div 0.2$ cm$^{-3}$  (soft ionizing spectrum).
Figure~15 shows that $\Upsilon_{\rm Si\,II} / \Upsilon_{\rm C\,II}$
is practically constant for both ionizing fields.
The behavior of
$\Upsilon_{\rm Fe\,II} / \Upsilon_{\rm C\,II}$ is, however, different: in
both cases we may expect to observe non-similar profiles of the
\ion{C}{2} and \ion{Fe}{2} lines. 

Thus we can conclude from these calculation that
independently on the background ionizing field the distribution of
\ion{C}{2} does trace the distribution of \ion{Si}{2} in diffuse
clouds (0.1 cm$^{-3} \la n_{\rm H} \la 10$ cm$^{-3}$)
and therefore we can use the unsaturated \ion{Si}{2} profiles to
measure $N$(\ion{C}{2}) with a sufficiently high accuracy. 
In the present case we were able to describe with the same velocity distribution
not only the \ion{C}{2} and \ion{Si}{2} lines, but profiles of the other
elements. This fact seems to justify the suggestion that the density fluctuations
in the \hh cloud are rather small and thus the derived $N$(\ion{C}{2}) value
can be considered as reliable.

\section{Discussion}

We turn now to the question of the physical conditions
in the \zabs =3.025 DLA. The measured column densities of
metals and \hh can be used to estimate the 
volumetric gas density and the intensity of the
UV radiation field.
We begin with the metal absorptions. Then we consider if
the observed \hh distribution over the excited rotational
states can be described with metals 
in a self-consistent way. 
After that we compare
the revealed [$\alpha$-element/iron-peak] ratio and
abundances with those known from different DLAs.

\subsection{Metal absorption from the ground and excited states} 

The absorption lines from 
the fine-structure levels of atoms and ions
provide comprehensive information about
the physical properties in the DLAs 
(e.g., Bahcall \& Wolf 1968; Silva \& Viegas 2001).

In equilibrium,
the population ratio of the upper level $n_2$ to the lower level 
$n_1$, in ions with a doublet fine structure in the ground state,
is given by
\begin{equation}
\frac{n_2}{n_1} \simeq \frac{Q_{1,2} + w_{1,2}}
{A_{2,1}}\; ,
\label{eq:E14c}
\end{equation}
where $w_{1,2}$ and $Q_{1,2}$ are
the photon absorption and the collisional
excitation rates, respectively,
and $A_{2,1}$ is the radiative transition probability.

In the present analysis, we are interested
mainly in the physical conditions in the gas where both the \hh
and the metal absorptions have been detected. We assume that
this absorbing region is homogeneous and therefore the
number density ratios are equal to the corresponding ratios
of column densities, $n_2/n_1 = N_2/N_1$. We neglect 
fluorescence since this indirect excitation process has usually
lower rates as compared to the direct collisional excitations
and/or the IR radiative absorption rates. The radiative decay and
the de-excitation rates are also ignored since 
in our case they have low efficiency. 

In the main component of the \zabs = 3.025 system 
the following quantities were
estimated from the data listed in Table~3:
$N$(\ion{H}{1}) = $(2.521\pm0.035)\times10^{20}$ cm$^{-2}$, 
$N$(\ion{C}{2}) = $(5.05\pm0.28)\times10^{15}$ cm$^{-2}$,
\ion{C}{2}$^\ast$/\ion{C}{2} = $(4.5\pm0.3)\times10^{-3}$,
\ion{C}{1}/\ion{C}{2}  $< 8\times10^{-5}$, 
\ion{O}{1}$^\ast$/\ion{O}{1} $< 3\times10^{-5}$, and
\ion{Si}{2}$^\ast$/\ion{Si}{2}  $< 2\times10^{-3}$. 

The ground state of the C$^+$ ion consists of two levels
$2s^2 2p$~$^2$P$^0_{1/2,3/2}$ with the energy difference of 
$\Delta E = 91.32$~K and the spontaneous decay rate of
$A_{2,1} = 2.291\times10^{-6}$ s$^{-1}$.
The excited J = 3/2 level may be
populated by collisions with electrons, e$^-$, hydrogen atoms, 
H$^0$, protons, H$^+$, and molecules, \hh. Among them the highest
rate is found for collisions with electrons~:
$Q^{{\rm e}^-}_{1,2} \simeq 2\times10^{-7}\,n_{\rm e}$\,
s$^{-1}$ in the range $10^2$~K $< T_{\rm kin} < 10^3$~K
(see, e.g., Figure~3 in Silva \& Viegas 2001).
But the electron density in \ion{H}{1} regions
comes mainly from
the photoionization of carbon, $n_{\rm e} \simeq
{\rm (C/H)}\,n_{\rm H}$,
which is approximately equal to $3\times10^{-5}\,n_{\rm H}$ for
the \zabs = 3.025 DLA. Then 
$Q^{{\rm e}^-}_{1,2} \simeq 6\times10^{-12}\,n_{\rm H}$\,
s$^{-1}$ 
becomes lower than the H$^0$--C$^+$ collisional rate 
for the same temperature interval where  $Q^{{\rm H}^0}_{1,2}$
is almost constant and equals
$Q^{{\rm H}^0}_{1,2} \simeq 1.45\times10^{-9}\,
n_{\rm H}$\, s$^{-1}$
(Launay \& Roueff 1977a).
The rate of the photon absorption in the \zabs = 3.025 DLA
is mainly induced by 
the cosmic microwave background radiation 
(Molaro et al. 2001b) and equals to $w_{1,2} \simeq 1.1\times10^{-9}$
s$^{-1}$.

From (\ref{eq:E14c}) and taking the limiting values
of the measured \ion{C}{2}$^\ast$/\ion{C}{2} ratio,
one obtains  
$n_{\rm H} \simeq 6\div7$ cm$^{-3}$.
The low limit of $n_{\rm H} \ga 4$ cm$^{-3}$ 
stems from the analysis of the \hh lines (see \S\,4.2).

To check the consistency of the estimated $n_{\rm H}$
with other metal absorption lines, we can use
the upper limits to the column densities of
\ion{O}{1}$^\ast$ and \ion{Si}{2}$^\ast$.
The equilibrium equation for the three-level system
($^3$P$_2$, $^3$P$_1$, $^3$P$_0$ states of O$^0$)
can be written as (cf. Molaro et al. 2001a)
\begin{equation}
\frac{n_1}{n_0} \simeq \frac{
(Q^{{\rm H}^0}_{0,1} + Q^{{\rm H}^0}_{0,2})}
{A_{1,0}}\; ,
\label{eq:E17}
\end{equation}
where the subscripts 0, 1, and 2 denote the ground (0) and 
excited (1,2) levels, respectively. The spontaneous decay rate 
$A_{1,0} = 8.95\times10^{-5}$ s$^{-1}$, and
the energy differences between the ground state and the
corresponding fine-structure levels are 
$\Delta E_{0,1} = 228$~K and $\Delta E_{0,2} = 326$~K. 

If  $n_{\rm e}$ is small, 
the rate coefficient for collisions with hydrogen
atoms is dominating. 
At the kinetic temperature $T_{\rm kin} = 400$~K, we find from
Launay \& Roueff (1977b)
$Q^{{\rm H}^0}_{0,1} = 
1.65\times10^{-11}\,n_{\rm H}$ s$^{-1}$ and
$Q^{{\rm H}^0}_{0,2} = 0.10\times10^{-11}\,n_{\rm H}$ 
s$^{-1}$.
Then (\ref{eq:E17}) yields
$n_{\rm H} < 150$ cm$^{-3}$, which is consistent
with the foregoing value.

The fine-structure level of the Si$^+$ ion 
($^3$P$^0_{3/2}$) has an effective excitation temperature
$\simeq 414$~K which means that
the estimated upper limit
to $N$(\ion{Si}{2}$^\ast$) is less sensitive to low gas densities
in this last case.
Besides, absorption features produced by \ion{Si}{2}$^\ast$ must come
only from \ion{H}{2} regions as argued by Spitzer \& Jenkins (1975).
The absence of the \ion{Si}{2}$^\ast$  absorption might indicate that
there is no dense \ion{H}{2} gas region on the line of sight. 
This suggestion
is also supported by the absence of the \ion{N}{2}$^\ast$ 
$\lambda1084.562$ \AA\, and $\lambda1084.580$ \AA\, lines  
in the \zabs = 3.025 system.
Both facts show that the \ion{C}{2}$^\ast$ and 
\ion{C}{2} absorptions
arise predominantly in a warm \ion{H}{1} gas in agreement with
our model described in \S\,3.6.

\subsection{\hh distribution over the rotational states}

We find no pronounced differences between the H$_2$ abundances
in the \zabs = 3.025 DLA system and in the Galactic 
diffuse clouds which have
comparable \ion{H}{1} column densities. 
The high excitation temperature in our system,
$T_{\rm ex} = 825\pm110$~K, is also consistent
with previous observations in the Galaxy. 
For instance, in a pioneer work
by Spitzer \& Cochran (1973) it has been shown that 
$T_{\rm ex} > 300$~K within
diffuse interstellar clouds where the \hh column density in the
J = 0 rotational level $N(0) < 10^{15}$ cm$^{-2}$,
and the temperature as high as 1100~K has been detected by
Morton \& Dinerstein (1976) toward $\zeta$~Pup.

It has been discussed in \S\,3.4  
that the populations
in the excited rotational states of \hh
cannot be supported only by collisions
in a gas whose kinetic temperature is as high as 800~K.
We calculated
the \hh collisional excitation rates
by using the analytical approximations given in
Shull \& Beckwith (1982).
The deduced rates are listed in Table~5, columns 3 and 4
for $T_{\rm kin} = 400$~K and 1000~K, respectively.
In columns 5 and 6 we show, for comparison, the rates
obtained in different models by Nishimura (1968) and Elitzur \&
Watson (1978). As noted by Shull \& Beckwith,
the accuracy of the collisional excitation rates
are not sufficiently high and different models may produce
deviations of about 50\%.
The employed analytical fit
to the H$^0$--\hh excitation rates agrees with
the Elitzur \& Watson calculations to better than 25~\%,
but gives much lower values than the Nishimura rates.
The radiative rates listed in column 2
were taken from Turner, Kirby-Docken, \& Dalgarno (1977).

The collisional excitation of the J = 3 level is not expected
to be effective in our case. Indeed,
the J = 2 level has longer radiative lifetime when compared with
the J = 3 level ($t_{{\rm J}=2} = 3.4\times10^{10}$ s\, and 
$t_{{\rm J}=3} = 2.1\times10^9$ s, respectively), 
and therefore the J = 2 level is
more sensitive to the collisional de-excitation. 
The critical density, 
$n^{\rm cr}_{\rm H} = A_{\rm 2 \rightarrow 0}/
q_{\rm 2 \rightarrow 0}$, 
above which collisional de-excitation becomes important is
$n^{\rm cr}_{\rm H} = 3.6$ cm$^{-3}$. 
Thus, $n_{\rm H} \ga 4$ cm$^{-3}$ is required to maintain
the observed $N(2)/N(0)$ ratio at $T_{\rm kin} = 400$~K.
The absence of a fracture at lower J in the graph
$\log [N({\rm J})/g({\rm J})]$ versus $E$(J), shown in Figure~8, can be
used to infer an upper limit for the gas density,
$n_{\rm H} \la 74$ cm$^{-3}$.
Both limits are consistent with $n_{\rm H}$
estimated from the measured $N$(\ion{C}{2}$^\ast$)/$N$(\ion{C}{2}) ratio.

Now we can set restrictions on the rate of photo-dissociation, $I$, 
in the \zabs = 3.025 molecular cloud
assuming that the \hh 
molecules are formed on grains with rate $R$ and
partly destroyed through the fluorescence process.
In equilibrium, the fractional molecular abundance
is given by (Jura 1974)
\begin{equation}
f_{{\rm H}_2} \simeq 2Rn_{\rm H}/I\; .
\label{eq:E19}
\end{equation}

To estimate the rate of formation of \hh, we use the 
approximation from Black \& Dalgarno (1977)
\begin{equation}
R \simeq 10^{-6}\,T^{0.5}_{\rm kin}\,n_{\rm H}\,n_{\rm d}\;
{\rm cm}^{3}~{\rm s}^{-1}\; .
\label{eq:E19a}
\end{equation}
Here the number density of dust grains, $n_{\rm d}$, is
related to the total hydrogen density, $n$, through
\begin{equation}
n_{\rm d} = 10^{-12}\,n = 10^{-12}\,
(n_{\rm H} + 2\,n_{{\rm H}_2})\; , 
\label{eq:E19b}
\end{equation}
which is the usual relation for the interstellar diffuse clouds.

Then, accounting for the dust-to-gas ratio $\tilde{k} \simeq 0.05$
(see \S\, 3.5) and the fact that $n_{{\rm H}_2} \ll
n_{\rm H}$ in the \zabs = 3.025 cloud, we find
\begin{equation}
R \simeq 10^{-18}\,T^{0.5}_{\rm kin}\,\tilde{k}\,n^2_{\rm H} \simeq
5\times10^{-20}\,T^{0.5}_{\rm kin}\,n^2_{\rm H}\; 
{\rm cm}^{3}~{\rm s}^{-1}\;.
\label{eq:E19c}
\end{equation}
Then for $T_{\rm kin} = 400$~K, \hh forms at a rate
$R \simeq 1.0\times10^{-18}\,n^2_{\rm H}$\, cm$^{3}$~s$^{-1}$.

This value of $R$ will be consistent 
with the upper limit found by Jura (1974) in the Galactic ISM,
$R_{\rm G} \la 2\times10^{-16}$\,
cm$^3$~s$^{-1}$,\, if $n_{\rm H} \la 14$ cm$^{-3}$ at \zabs = 3.025. 
By this we confine $n_{\rm H}$ in the interval
4 cm$^{-3}$ $\la n_{\rm H} \la 14$ cm$^{-3}$ which provides
a self-consistency between the observed $N(2)/N(0)$ ratio
and the maximum formation rate of \hh under the condition
that the physical properties of the dust grains in the Milky
Way and at \zabs = 3.025 are similar.

Since the molecular hydrogen fractional
abundance in the main component of the \zabs = 3.025 absorber
is $f_{{\rm H}_2} = (3.3\pm0.2)\times10^{-6}$, cloud
parameters $n_{\rm H} \simeq 6$ cm$^{-3}$ and $T_{\rm kin} =
400$~K will yield $I \simeq 2\times10^{-10}$ s$^{-1}$
which agrees well with the values measured in the
interstellar clouds containing optically thin \hh:
$0.5\times10^{-10}$ s$^{-1}$ $\leq I_{\rm G} \leq
16.0\times10^{-10}$ s$^{-1}$
(Jura 1975a). Thus, the radiation field in the \zabs = 3.025
cloud is quite comparable with those measured in 
transparent interstellar clouds in the MW.

We now evaluate the total hydrogen 
density, $n_{\rm H}$, the photo-absorption rate,
$\beta_0$, and the steady state production rate
of molecular hydrogen, $R$, 
from the relative population of the
different rotational levels of \hh.
Our analysis is, in general, the same as described
by Spitzer \& Zweibel (1974) and
by Jura (1974, 1975a,b). However, we re-calculated the
formation efficiency coefficients 
$\varepsilon_4$ and $\varepsilon_5$
for the J = 4 and 5 levels
including upper rotational levels with a cutoff
above J$_{\rm max} = 15$.
These coefficients were calculated from
the Boltzmann distribution of molecules formed upon grain surfaces
with an effective temperature, $T_{\rm f}$ (for details, see Black \&
Dalgarno 1976).
Since a newly formed molecule quickly cascades to the 
ground vibrational level, we assume that the newly formed
molecules populate the 
(v=0, J=4) and (v=0, J=5) levels.
It is usually assumed that molecules are formed at
$T_{\rm f} \simeq 17400$~K, which is 
the temperature equivalent of the binding energy
of \hh distributed statistically among the rotation-vibration
levels upon formation (e.g. Spitzer \& Zweibel 1974).
Then we find
$\varepsilon_4 = 0.21$ and
$\varepsilon_5 = 0.65$ (cf. $\varepsilon_4 = 0.19$
and $\varepsilon_5 = 0.44$ from Jura 1975b).

The concentrations of \hh in the J = 4 and 5 levels may be
approximately estimated from the statistical equilibrium equations
\begin{equation}
A_{\rm 4+K,2+K}\,n_{\rm 4+K} = 
\varepsilon_{\rm 4+K}\,R\,n^2_{\rm H} +
\beta_0\sum_{\rm J'}\,
p_{\rm 4+K,J'}\,n_{\rm J'}\; , 
\label{eq:E22}
\end{equation}
where ${\rm J}' = K,K+2,K+4$, and
$K = 0$ or 1 for J = 4 or 5, respectively.

In these equations, the pumping efficiency coefficients
$p_{i,j}$ are taken from Jura (1975a). The quantity
$p_{i,j}$ shows the probability (averaged over the
different Lyman and Werner bands) that photo-absorption
from the $j$th rotational level will lead to the
population of the $i$th rotational level in the ground
v = 0 vibrational state. 

From (\ref{eq:E22}) we can calculate $R\,n_{\rm H}$
using the measured $N(4)/N$(H) and $N(5)/N$(H) ratios.
It gives $R\,n_{\rm H} \simeq 3\times10^{-16}$ s$^{-1}$.
If we take $R \simeq 5\times10^{-17}$ cm$^3$~s$^{-1}$
from Spitzer \& Cochran (1973), then
$n_{\rm H} \simeq 6$ cm$^{-3}$. Now the \hh destruction rate
can be estimated from (\ref{eq:E19}) which gives again
$I \simeq 2\times10^{-10}$ s$^{-1}$ as in the foregoing estimations.
If $I \simeq 0.11\beta_0$  (Jura 1974), then the rate
of photo-absorption is 
$\beta_0 \simeq 2\times10^{-9}$ s$^{-1}$. On the other hand,
the average $\beta_0$ for the intergalactic UV background at 
$z \simeq 3$ is about $2\times10^{-12}$ s$^{-1}$ (Srianand \& 
Petitjean 1998), while the MW interstellar radiation field value
ranges between
$5\times10^{-10}$ s$^{-1} \leq \beta_0 \leq 16\times10^{-9}$ s$^{-1}$
(Jura 1975a). 
Thus, our measurements show a close similarity with the average
characteristics of the UV field found in the MW.

For the neutral hydrogen column density
$N$(\ion{H}{1}) $\simeq 2.5\times10^{20}$ cm$^{-2}$
and $n_{\rm H} \simeq 6$ cm$^{-3}$,
the linear size of the \hh-bearing cloud
along the line of sight is equal to
$D \simeq 14$ pc which yields
a mass of $M \simeq 0.5D^3n_{\rm H}m_{\rm H} \simeq 200M_\odot$, 
assuming spherical geometry.

\subsection{[$\alpha$-element/iron-peak] ratios}

The elemental column densities obtained 
following the procedure described
in sections 3.5 and 3.6 together with the 
relative abundances are summarized in Table 4.

\noindent
Among the elements belonging to the iron-peak group, Zn  
is a good indicator of the absolute metal abundance
in the system, since Zn is known 
to be  essentially undepleted in the ISM 
(Pettini et al. 1999). Measurements of Zn in high redshift systems 
have been proven to be very challenging since
the \ion{Zn}{2} resonance lines fall at wavelengths 
$>$ 8000 \AA\, where spectrograph and CCD
efficiency decrease while sky emissivity increases.
The Zn measurement presented here and that 
in the DLA at \zabs = 3.39 toward Q0000--2621 are the only  
ones  obtained for systems with $z \ge 3$.
The derived [Zn/H] = $-1.50\pm0.05$ is in line with the bulk of 
[Zn/H] measurements found in the redshift interval 
$2 < z < 2.5$ although higher by 0.6 dex than that of 
the DLA at \zabs = 3.39 toward Q0000--2621 confirming 
the presence of significant spread in absolute abundance
in DLA with comparable redshifts. The straight mean
of the Zn measurements in the redshift 
interval from $2 \le z \le 3$ is [Zn/H] $\approx -1.2$, 
and does not differ from the column density weighted mean 
for  the same redshift interval (Vladilo et al. 2000).
This new measurement therefore supports the general metallicity
behavior, characterized by the intrinsic spread for 
a given redshift and by a mild trend
of increasing abundance with the decreasing redshift 
if not a lack of evolution when
the hydrogen column density weighted mean is considered.

As described in previous \S\, 3.6, our model keeps 
the constant ratio between the matching components. 
With this condition imposed, we 
cannot account for different chemical abundances
or the presence of
different dust pattern depletion among the components 
of the system.
It may be possible that a contribution from warmer gas or
different dust depletion is responsible for the observed 
difference in  profile shapes of \ion{Fe}{2} and other ions.

Only recently it became possible to measure oxygen with an acceptable
accuracy using  the  fainter \ion{O}{1} lines falling in the 
Ly$\alpha$ forest instead of the always strongly saturated 
\ion{O}{1}\,$\lambda1302$ \AA\, transition 
(Lopez et al. 1999;
Molaro et al. 2000; Dessauges-Zavadsky et al. 2001). 
In the present paper, the  O abundance is 
derived from the set of eleven OI lines  
used either for direct fitting
or just for the consistency test.
  
The [O/Zn] is the most informative ratio to trace back
the kind of chemical evolution of the absorber. Both elements 
show little affinity with
dust grains and are good representatives of 
$\alpha$-chain and iron-peak elements, respectively. 
If the system is in the early stages of 
chemical evolution,
the relative abundances should reflect the yields of 
Type II SNe characterized
by the enhancement of $\alpha$-elements. 
In fact, it was rather surprising 
to find little evidence for this enhancement when elements such 
as S and Si, once corrected by dust, where
used in combination with Zn 
(Centuri\'on et al. 2000; Vladilo 1998) and the ratio [O/Zn] = 0.1
has been measured in the DLA toward 
Q0000--2621 (Molaro et al. 2001a).

The ratio [O/Zn] = $0.68\pm0.08$, found here, represents 
the first clear cut evidence for a genuine  enhancement of the
$\alpha$-elements comparable to that observed in the metal poor 
stellar population of the Milky Way (Israelian et al. 2001). 
The oxygen result is also supported by the other 
$\alpha$-elements we have been able to measure, 
namely [Ar/Zn] = $0.34\pm0.07$,  
[Si/Zn] = $0.56\pm0.06$, 
and [S/Zn] = $0.41\pm0.08$ (if we take $\log N$(\ion{S}{2}) =
$14.73\pm0.01$ cm$^{-2}$ from Prochaska \& Wolfe 1999).
Argon abundance is slightly below the values obtained from the
other $\alpha$-elements, but this is not 
surprising since Ar is sensitive to photoionization effects
and shows in the Galaxy even larger
depletion factors (Sofia \& Jenkins 1998).

All results available at present allow us to conclude that
the DLAs probably comprise a mix of chemical 
abundance pattern with both low and {\it standard} 
$\alpha$-element enhancement. 
It is not clear at the moment if this reflects
the different enrichment histories or just the same type 
of galaxies caught at different stages of their chemical evolution.
It is interesting to note that estimated in the H$_2$-bearing cloud 
the ratio ${\rm (Ar/O)} = -2.65$  
is comparable to that observed in the DLA toward Q0000--2621, 
(Ar/O) = $-2.25$ (Molaro et al. 2001a),
in particular when
allowing for a small degree of Ar photoionization. 
As argued  by Henry \& Worthey (1999)
the similarity of these ratios in a variety of 
galactic systems from the MW to Blue Compact galaxies 
strengthens the case for the existence of a universal IFM.

It is commonly assumed that nitrogen is mostly produced by the
intermediate massive stars (Henry, Edmunds, \& K\"oppen  2000). 
In DLAs,  N shows a  scatter well in excess of observational 
uncertainties which has been interpreted as being due to the 
delayed release of N compared to O
(Centuri\'on et al. 1998).
Our value in the `[O/H]--[N/O]' 
plane has coordinates [O/H] = $-0.82$ and [N/O] = $-0.84$ and 
falls close to the line for a pure secondary
origin for N (cf. Figure~8 in Centuri\'on et al. 1998). 
If intermediate massive stars had not yet time 
to evolve and to contribute to the N enrichment then
this sets an upper limit of  
approximately 10$^8$ yr to the evolutionary age
of the \zabs = 3.025 system. The young age of the system under study
is also consistent with the observed $\alpha$-element enhancement.
 
Phosphorus has been measured in the DLA toward Q0000--2621 
(Molaro et al. 2001a) and toward Q1759+7539 (Outram et al. 2000). 
P cannot be measured in halo 
stars and thus the damped Ly$\alpha$ systems provide
a unique astrophysical site where this element can be measured in
low metallicity environments.  Our determination is based 
only on the \ion{P}{2}\,$\lambda963.8$ \AA\, line 
and may be slightly overestimated due to \ion{H}{1} contamination. 
We have here the mean value of [P/Si] $\approx -0.2$ 
which suggests the presence of a moderately odd-Z/even-Z 
effect which is a common feature in all P measurements in DLAs so far.

\section{Conclusions}

The main results of this paper are as follows.

(1) In the high resolution UVES spectrum of the quasar 0347--3819
we have identified over 80 absorption features with the Lyman
and Werner lines arising from the J = 0 to 5 rotational levels
of the ground electronic-vibrational state of \hh. These lines
belong to the \zabs = 3.025 damped Ly-$\alpha$ system.

(2) The application of the standard Voigt fitting analysis to the
main subcomponent at $z_{{\rm H}_2} = 3.024855\pm0.000005$
gives the total \hh column density of
$N$(H$_2$) = $(4.10\pm0.21)\times10^{14}$ cm$^{-2}$ and
the Doppler width $b_{{\rm H}_2} = 2.80\pm0.45$ km~s$^{-1}$.
The kinetic gas temperature is less than 430~K.

(3) The fractional abundance of \hh in the main component
is equal to $f_{{\rm H}_2} = (3.3\pm0.2)\times10^{-6}$, while
being compared to the total \ion{H}{1} column density
it yields $f_{{\rm H}_2} = 
(1.94\pm0.10)\times10^{-6}$. These values are similar to those
observed in the ISM diffuse clouds with low color
excesses, $E(B-V) < 0.1$.
Using the total hydrogen column density in the main component
$N$(\ion{H}{1}) $\simeq 2.5\times10^{20}$ cm$^{-2}$ and
the approximate relationship $N$(H) $\simeq 
7.5\times10^{21}\,E(B-V)$ cm$^{-2}$~mag$^{-1}$, which
assumes that the gas and dust grains are uniformly
mixed (Spitzer \& Jenkins 1975), 
we obtain $E(B-V) \simeq 0.03$ mag
in the H$_2$-bearing cloud.
For the second strong component with $N$(\ion{H}{1}) $\simeq
1.5\times10^{20}$ cm$^{-2}$ 
at $\Delta v \simeq -17$ km~s$^{-1}$ we set a very low limit to 
$f_{{\rm H}_2} < 3.3\times10^{-9}$ ($3\sigma$). 
The \zabs = 3.025 system is not a homogeneous cloud
but rather a mixture of cold and warm gas.

(4) The total column densities in the ortho- and para-\hh states
yield the ratio $N_{\rm ortho}/N_{\rm para} = 3.1\pm0.3$
which is close to 3:1 expected from the hot \hh
formation upon grain surfaces. 
Similar ratios of $2.4\pm0.9$ and $3.1\pm0.8$
are estimated at \zabs = 2.33771 and 2.8112 toward
Q1232+0815 and Q0528--250 from the published data
by Srianand et al. (2001) and Srianand \& Petitjean (1998),
respectively.
These ratios are, however, higher
than the `freeze-out' ortho:para-\hh = 1:4 predicted for the
early universe at $z < 20$ (Flower et al. 2000). It appears that
\hh detected in these three DLAs has undergone a re-forming phase
in the period between $z \simeq 20$ and $z \simeq 3$
($\Delta t \simeq 2$ Gyr).

(5) We found that the population of the 
low rotational levels can be represented by
a single excitation temperature of $T_{\rm ex} =
825\pm110$~K. Such high $T_{\rm ex}$ agrees well with the results by
Spitzer \& Cochran (1973) who showed that $T_{\rm ex} > 300$~K
within diffuse clouds 
which have $N({\rm J=0}) < 10^{15}$ cm$^{-2}$,
as it is in our case.

(6) From the measured [Cr/H] and [Zn/H] ratios we calculated
the gas-to-dust ratio $\tilde{k} = 0.032\pm0.005$ in the
whole \zabs = 3.025 system 
(the ratio [Cr/Zn]
is usually used to indicate the presence of dust in DLAs).
The derived \hh fractional abundance
fits well in the linear relation between $\log f_{{\rm H}_2}$
and [Cr/Zn] supporting our previously obtained results for
other DLA systems (Levshakov et al. 2000b). 

(7) From the observed
population ratios of \hh in different rotational states and
the \ion{C}{2}$^\ast$/\ion{C}{2} ratio we infered the hydrogen
number density $n_{\rm H}$ in the H$_2$-bearing cloud
$n_{\rm H} \simeq 6$~cm$^{-3}$,
and the cloud dimension along the line of sight
$D \simeq 14$ pc which gives $M \simeq 200 M_\odot$.

(8) From the relative populations of \hh in the J = 4 and 5
rotational levels we
estimated the rate of
photo-destruction $I \simeq 2\times10^{-10}$ s$^{-1}$.
The photo-absorption rate, $\beta_0 \simeq I/0.11$, is thus
equal to $\beta_0 \simeq 2\times10^{-9}$ s$^{-1}$.
Taking into account that the MW interstellar radiation field value
ranges in
$5\times10^{-10}$ s$^{-1}$ $\leq \beta_0 \leq 16\times10^{-9}$ s$^{-1}$
(Jura 1975a),
we conclude that the UV radiation fields in the
\zabs = 3.025 absorbing cloud and
in the Galactic ISM are very much alike.
We also found that the formation rate of \hh
upon grain surfaces, $R\,n_{\rm H} \simeq 3\times10^{-16}$ s$^{-1}$,
is similar to that found in the Galaxy by Jura.

(9) The metal abundances from the \zabs = 3.025 DLA reveal
a pronounced [$\alpha$-element/iron-peak] enhancement with
[O,Si/Zn] = $0.6\pm0.1$ at the $6\sigma$ confidence level,
the first time that this abundance pattern is
unambiguously identified in a DLA system.
The same order of magnitude enrichment of $\alpha$-chain
elements has been observed in Galactic metal-poor stars.
The measured [N/O] ratio implies the chemical history age of this
DLA of $\la 10^8$ yr.

(10) The analysis of the metal profiles combined with the
H+D Lyman series lines yields a new estimation of the
hydrogen isotopic ratio
D/H = $(3.75\pm0.25)\times10^{-5}$ in this DLA.
It implies for the baryon-to-photon ratio, $\eta$, a value
in the interval $4.37\times10^{-10} \la \eta \la
5.32\times10^{-10}$, and 
the present-day baryon density
$\Omega_{\rm b}\,h^2 
\simeq 0.016 \div 0.020$, which is in good agreement with
a new analysis of the BOOMERANG experiment yielding
$\Omega_{\rm b}\,h^2 = 0.021\pm0.003$ (Netterfield et al. 2001).

\acknowledgments

S.A.L. gratefully acknowledges the hospitality of the European
Southern Observatory (Garching), Osservatorio Astronomico
di Trieste, and the National Astronomical Observatory
of Japan (Mitaka) where this work was performed. 
We also thank A. Wolfe \& J. X. Prochaska for making available the
calibrated Keck/HIRES spectrum of Q0347--3819,
and I. I. Agafonova for many valuable comments on data analysis.
We are greatly indebted to our referee for noting a shortcoming 
in the pipeline software and for helpful comments. 
The work of S.A.L. is supported in part by the 
RFBR grant No.~00-02-16007.

\appendix

\section{Correlated measurements}

In order to evaluate the uncertainties in the derived physical
parameters, one must know the standard errors, $\sigma_j$, for all
the data points. These errors, computed through the UVES data reduction
pipeline (Ballester et al. 2000) are not, however, statistically
independent. The pipeline data resampling to constant wavelength bins
introduces a correlation between the data point values. 
Besides the resampling procedure can change statistical properties
of the noise (e.g., the simplest 
linear interpolation scheme smooths the
random noise since the uncertainty of the data point value half way
between two measurements is less than that for the real detector
amplitudes at the limiting points).

Data point 
correlation is clearly seen in Figure~7 where the sequences of
points at the continuum level (e.g., panels L2-0R0 or L2-0P3)
show that neighboring points differ from each other by less than
$\sigma_j$
(for the data shown in Fugure~7, the correlation coefficient 
$r \simeq +0.77$). 
The correlation length, $\xi$, of about 1.5 bin can
be deduced from the analysis of the autocorrelation function
(ACF)\footnote{\,The ACF is defined, e.g., in Bendat \& Piersol (1971)}
of the intensity fluctuations in the continuum windows free from
absorption-line features. This ACF is
shown by histogram in Figure~16a where an exponential correlation
function $f(\Delta n) = \exp (-\Delta n/\xi)$ of the simplest
data model, --
the first order autoregression (the Markov process), -- is overdrawn
by the smooth curve (here $\Delta n$ is the scale shift).

The correlated data make the standard
$\Delta\chi^2$ calculation of parameter confidence intervals questionable.
To investigate this point we performed the following Monte Carlo (MC)
analysis in combination with the least-squares method.

It is well known that the least-squares method (LSM) is
computationally equivalent to the $\chi^2$-minimization 
(e.g., Meyer 1975). 
It is also known that the LSM is an unbiased and minimum variance
estimator (Gauss-Markov theorem) irrespective of the distribution
of the measurements including correlated data as well (Aitken 1934).
Therefore, in the MC analysis, we used the best fitting parameters
(the parameter vector $\hat{\theta}$, see \S\,3.6)
found by means of the standard $\chi^2$-minimization, and calculated 
uncertainties of the $\hat{\theta}$-components
in the vicinity of the global minimum of the objective function
by using the MC procedure.

Our computational MC procedure consists of the following steps:

\smallskip\noindent
(1)\, A simulation box in the parameter space is specified by
fixing the parameter boundaries
$\hat{\theta}_i(1-\beta) < \theta_i < \hat{\theta}_i(1+\beta)$;
in our case $\beta = 0.3$.
  
\smallskip\noindent
(2)\, Initial guess
$\theta^{(0)}$ is chosen arbitrary within the simulation box.

\smallskip\noindent
(3)\, A random realization of the `observational' points is generated
in the followin way: the 
$j$th data point value 
is altered with a probability $P$ by means of 
the relation
$$
{\cal F}^{\rm rnd}_j = {\cal F}^{\rm obs}_j + \sigma_j\,g\,,
$$
where $g$ is a random normal variable with zero mean and unit 
dispersion, and ${\cal F}^{\rm obs}_j$ is the normalized observational
intensity\footnote{\, This randomization decreases the
correlation between consecutive bins, but increases the noise. 
For instance, if
$P = 0.7$ the correlation length equals $\xi \simeq 1.0$ bin  
(see Figure~16b) but the noise level is about a factor of 
$\sqrt{2}$\, higher.}
in the $j$th bin ($j = 1,2,\ldots,{\cal N}$; with ${\cal N}$ 
being the total number of data points).

\smallskip\noindent
(4)\, Objective function is minimized and all solutions
with $\chi^2 < 1.10$ are collected. 

\smallskip\noindent
(5)\, The whole procedure is repeated many times to collect a 
representative sample (i.e. the number of measurements
${\cal K} > 30$). Then 
the sample mean and standard deviation are calculated.

To compare both $\Delta\chi^2$ and MC 
results we carried out the MC procedure for
the $b$ parameter and the \hh abundances at different J
levels. 
Nineteen \hh lines 
(L3-0R0, L7-0R0, L1-0P1, L3-0P1, L14-0P1, W0-0R2, W0-0Q2, L3-0R2,
L4-0R2, L5-0R2, L2-0R3, L2-0P3, L3-0P3, L4-0R3, L15-0R3, L5-0R4,
L8-0P4, L7-0P5, L12-0R5; see Figs.~5 and 6)
have been chosen and fitted simultaneously 
with the individual Gaussian profiles.
The obtained estimations
(for ${\cal N} = 177$ and ${\cal K} = 181$)
summarized in Table~6 show that in this case $\Delta\chi^2$
yields practically the same mean values but systematically
higher standard deviations. 
Since statistical properties of the noise
are not exactly known we use in the present paper
the most conservative $\Delta\chi^2$
estimations of the standard deviations and of the mean
values. 

\section{Metal systems toward QSO 0347--3819}

We have thoroughly investigated all possible coincidences of H$_2$,
\ion{H}{1}, \ion{D}{1}, and metal absorption features from the
\zabs = 3.025 damped Ly-$\alpha$ system with metal absorptions from
other systems observed toward Q0347--3819. Positions of the found
blends are indicated in Figures ~5, 6, 10, and 11. The strongest systems
detected in the spectral ranges $\lambda\lambda = 3680 - 4880$ \AA,
$4880 - 6730$ \AA\, (this portion of the spectrum obatined at the
Keck telescope was kindly shared with us by Wolfe \& Prochaska), 
$6730 - 8520$ \AA,\, and $8685 - 10000$ \AA\, are listed below.
Since almost all identified metal profiles exhibit complex structures,
the quoted redshifts are the ones of the strongest components or the
average redshifts in the cases when the components are not well separated.
In cases where the hydrogen Ly-$\alpha$ line lies outside the 
observable spectral range,
we consider a doublet of absorption lines caused by the same ion as an
individual metal system if the lines show identical profile shapes.
The identifications made in the \zabs = 3.025 DLA are not included in
the list. Possible blends are shown in parenthesis.

\bigskip\noindent
{\small
$z_1 =0.000019$\hspace{0.5cm}\ion{Ca}{2}\, 3934,3969(+W1-0R2)\\ 
$z_2 =1.4579$\hspace{0.9cm}\ion{Mg}{1}\, 2852; \ion{Mg}{2}\, 2796,2803;
\ion{Al}{2}\, 1670; \ion{Al}{3}\, 1854; \ion{Si}{2}\, 1526,1808;\\
\hspace*{2.8cm}\ion{Fe}{2}\, 1608,2260,2344,2374,2382,2586,2600\\
$z_3 =1.5159$\hspace{0.9cm}\ion{C}{4}\, 1548,1550\\
$z_4 =1.5263$\hspace{0.9cm}\ion{Mg}{2}\, 2796,2803;
\ion{Si}{2}\, 1526(+Ly-$\gamma\, z_{15}$),1808; 
\ion{Fe}{2}\, 2344,2382,2600\\
$z_5 =1.6768$\hspace{0.9cm}\ion{C}{4}\, 1548,1550\\
$z_6 =2.2140$\hspace{0.9cm}Ly-$\alpha$; \ion{C}{2}\, 1334; \ion{C}{4}\,
1548,1550; \ion{O}{1}\, 1302; \ion{Mg}{2}\, 2796,2803; \ion{Al}{2}\, 1670;\\
\hspace*{2.8cm}\ion{Al}{3}\, 1854; \ion{Si}{2}\, 1190(+W3-0Q3),1193,1260; 
\ion{Si}{3}\, 1206; \ion{Si}{4}\, 1393,1402;\\
\hspace*{2.8cm}\ion{Fe}{2}\, 1608,2344,2374,2382,2586,2600\\
$z_7 =2.3932$\hspace{0.9cm}Ly-$\alpha$(+Ly-$\beta$ DLA); \ion{C}{4}\, 1548,1550\\
$z_8 =2.4344$\hspace{0.9cm}Ly-$\alpha$; \ion{C}{4}\, 1548,1550\\
$z_9 =2.5371$\hspace{0.9cm}Ly-$\alpha$; \ion{C}{2}\, 1334; \ion{C}{4}\, 1548,1550;
\ion{Al}{2}\, 1670; \ion{Si}{2}\, 1260; \ion{Si}{4}\, 1393\\
$z_{10}=2.6504$\hspace{0.8cm}Ly-$\alpha$; \ion{C}{4}\, 1548,1550;
\ion{Si}{3}\, 1206; \ion{Si}{4}\, 1393,1402\\
$z_{11}=2.6526$\hspace{0.8cm}Ly-$\alpha$; \ion{C}{4}\, 1548,1550;
\ion{Si}{3}\, 1206; \ion{Si}{4}\, 1393,1402\\
$z_{12}=2.8102$\hspace{0.8cm}Ly-$\alpha$; \ion{C}{2}\, 1334; 
\ion{Si}{2}\, 1190,1193,1260(+\ion{Si}{2} 1193 DLA); \ion{Si}{4}\, 1393,1402\\
$z_{13}=2.8483$\hspace{0.8cm}Ly-$\alpha$, Ly-$\beta$(+Ly-$\gamma\, z_{19}$); 
\ion{C}{3}\, 977; \ion{C}{4}\, 1548,1550\\
$z_{14}=2.8992$\hspace{0.8cm}Ly-$\alpha$, Ly-$\beta$, Ly-$\gamma$, 
Ly-$\delta$; 
\ion{C}{4}\, 1548,1550; \ion{C}{3}\, 977;\\ 
\hspace*{2.9cm}\ion{Si}{3}\, 1206; \ion{Si}{4}\, 1393,1402\\
$z_{15}=2.9617$\hspace{0.8cm}Ly-$\alpha$, Ly-$\beta$, Ly-$\gamma$, Ly-$\delta$; 
\ion{C}{3}\, 977; \ion{C}{4}\, 1548,1550(+\ion{Si}{2} 1526 DLA);\\
\hspace*{2.9cm}\ion{N}{2}\, 1083; \ion{N}{3}\, 989; \ion{Al}{3}\, 1854; 
\ion{Si}{2}\, 1190; \ion{Si}{3}\, 1206\\
$z_{16}=2.9659$\hspace{0.8cm}Ly-$\alpha$, Ly-$\beta$, 
Ly-$\gamma$(+\ion{Si}{2} 1526 $z_4$), Ly-$\delta$, Ly-$\varepsilon$; 
\ion{C}{3}\, 977; \ion{C}{4}\, 1548;\\ 
\hspace*{2.9cm}\ion{N}{3}\, 989(+L11-0P2); \ion{Si}{3}\, 1206\\
$z_{17}=2.9792$\hspace{0.8cm}Ly-$\alpha$, Ly-$\beta$, Ly-$\gamma$; 
\ion{C}{2}\, 1036(+Ly-$\beta$ DLA); \ion{C}{3}\, 977; \ion{C}{4}\, 1548,1550; \\ 
$z_{18}=3.0217$\hspace{0.8cm}Ly-$\alpha$(+Ly-$\alpha$ DLA); 
\ion{C}{3}\, 977(+\ion{O}{1} 976 DLA); \ion{C}{4}\, 1548,1550;
\ion{N}{2}\, 1083; \\
\hspace*{2.9cm}\ion{N}{3}\, 989; \ion{Si}{4}\, 1393,1402\\
$z_{19}=3.0632$\hspace{0.8cm}Ly-$\alpha$, Ly-$\beta$, 
Ly-$\gamma$(+Ly-$\beta\, z_{13}$), Ly-$\delta$(+\ion{N}{3} 989 $z_{14}$); 
\ion{C}{3}\, 977; \ion{C}{4}\, 1548,1550
}

\clearpage

\begin{figure}
\figcaption[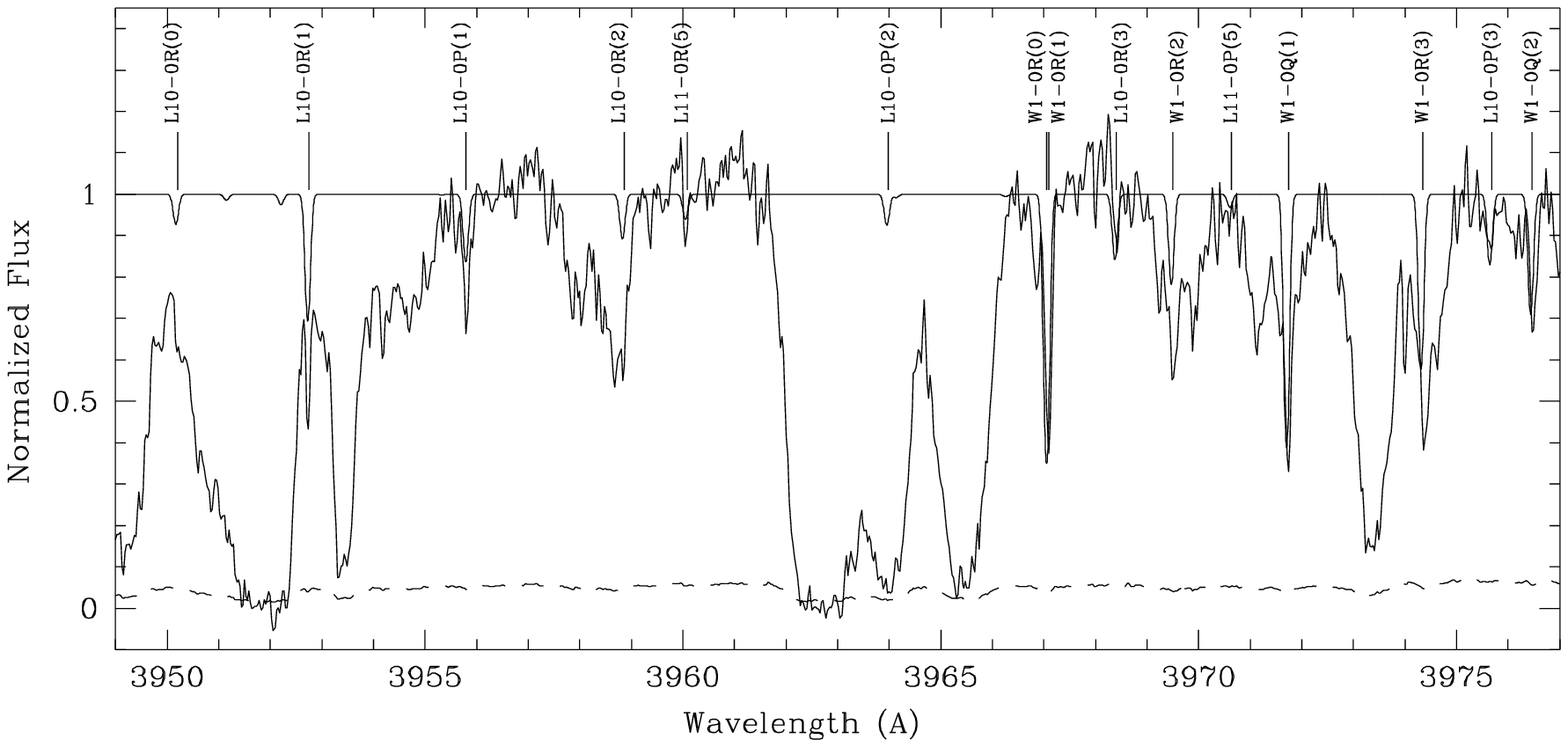]{
A portion of the normalized spectrum of Q0347--3819 in
the region of a few \hh lines from the Lyman and Werner bands
(labels `L' and `W', respectively) identified in the \zabs = 3.025
damped Ly$\alpha$ system. The noise level is indicated by the 
dashed curve
at the bottom of the panel. The general continuum was estimated by
using a polynomial fit to the `continuum windows' 
in all spectral range spanning about 1200 \AA. In cases where the
curvature of the QSO continuum at $z = 3.025$ is strong, the 
uncertainty in the local continuum level may be dominated by
uncertainties in the fitting procedure.
\label{fig1}}
\end{figure}

\begin{figure}
\figcaption[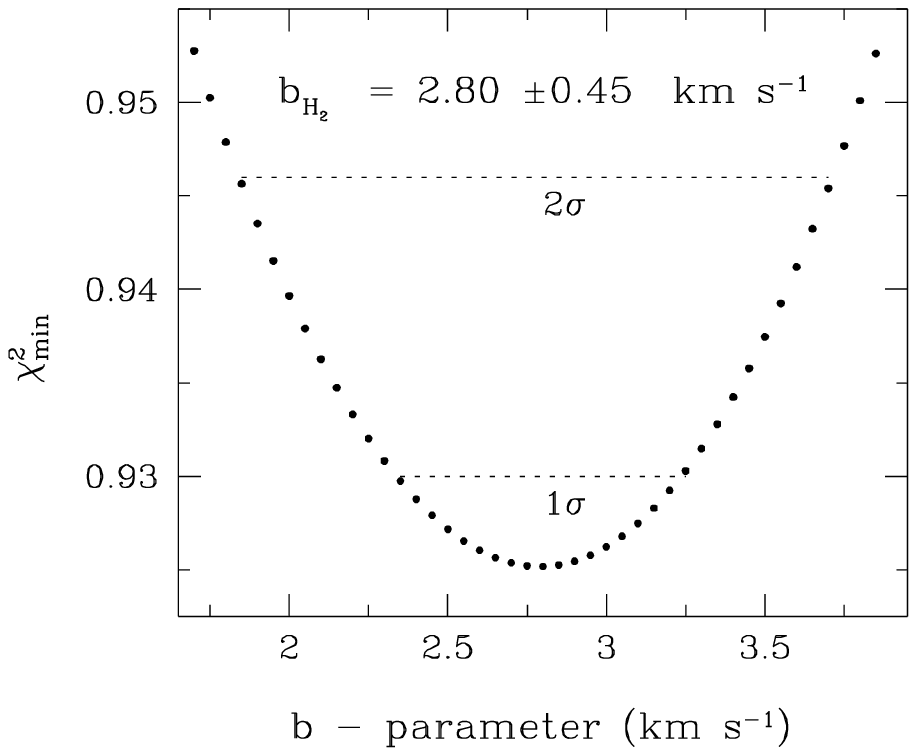]{
Confidence regions in the `$\chi^2 - b$' plane 
calculated from the simultaneous fit of the \hh lines from
Table~1 as described in the text.
The parabola vertex corresponds to the most probable value
of $b = 2.80$ km~s$^{-1}$.
\label{fig2}}
\end{figure}

\begin{figure}
\figcaption[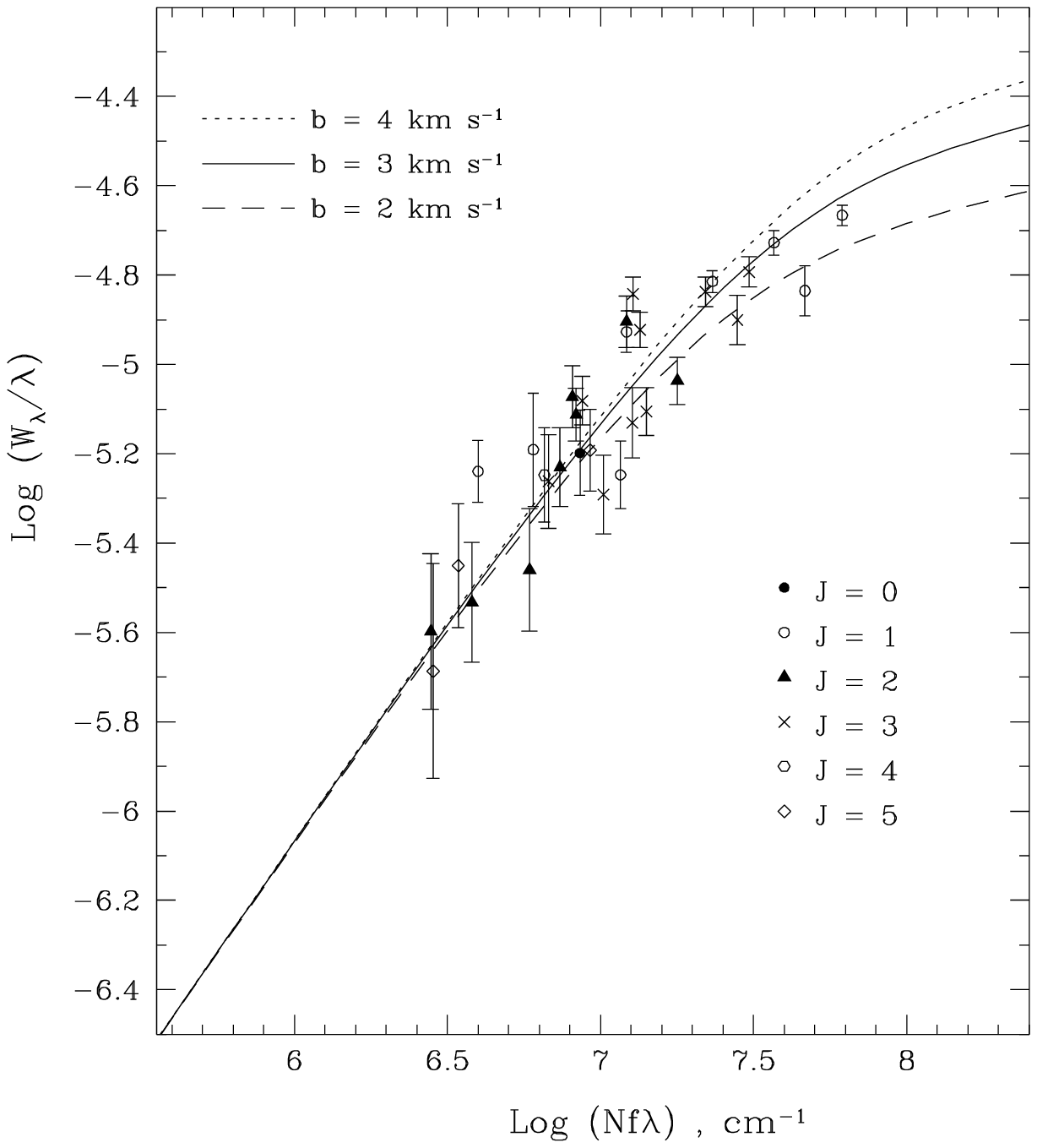]{
Curves of growth for 31 \hh lines (J = 0, 1, 2, 3, 4, and 5; 
see Table~1)
in the \zabs = 3.025 molecular cloud toward Q0347--3819. Most
probable Doppler parameter is $b_{{\rm H}_2} = 3$ km~s$^{-1}$. 
Curves with $b = 2$ km~s$^{-1}$ and 4 km~s$^{-1}$ restrict the 
$1\sigma$ confidence region for $b_{{\rm H}_2}$.
\label{fig3}}
\end{figure}

\begin{figure}
\figcaption[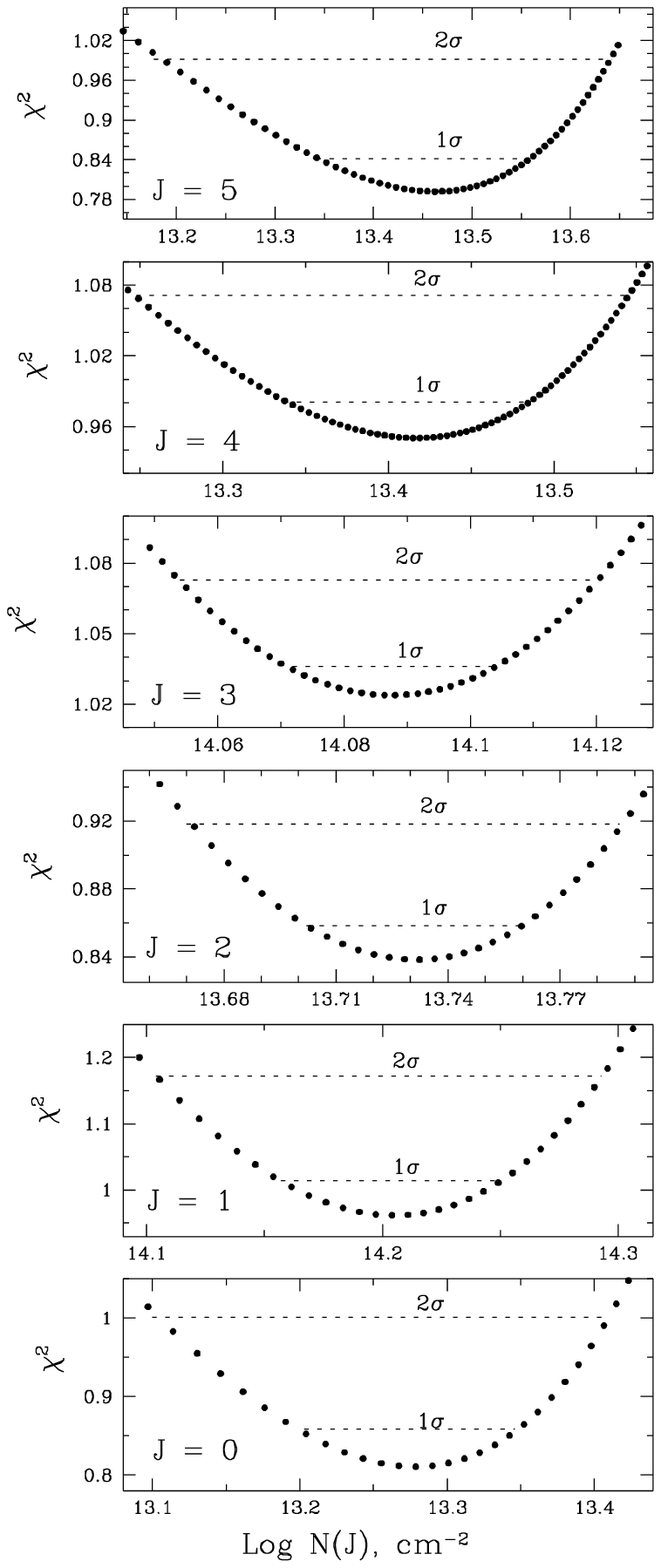]{
Confidence regions in the `$\chi^2 - {\rm Log} N({\rm J})$' planes 
calculated from the simultaneous fits of the \hh lines 
arising from the same J levels.
The parabola vertexes correspond to the most probable values of 
$N(0) = 1.90\times10^{13}$ cm$^{-2}$,
$N(1) = 1.60\times10^{14}$ cm$^{-2}$,
$N(2) = 5.40\times10^{13}$ cm$^{-2}$,
$N(3) = 1.22\times10^{14}$ cm$^{-2}$,
$N(4) = 2.60\times10^{13}$ cm$^{-2}$, and
$N(5) = 2.90\times10^{13}$ cm$^{-2}$.
\label{fig4}}
\end{figure}

\begin{figure}
\figcaption[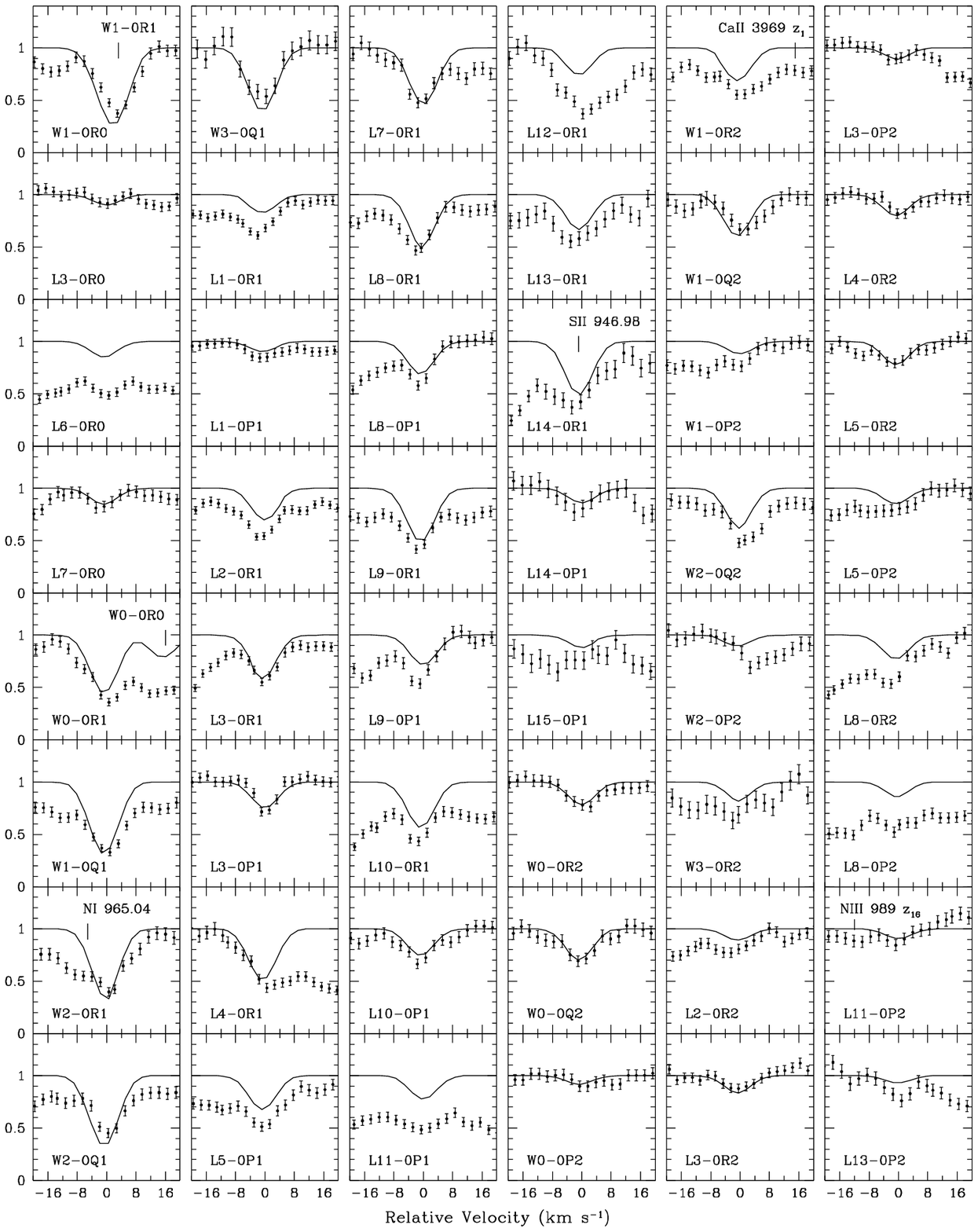]{
\hh absorption features associated with the \zabs = 3.025 damped
Ly$\alpha$ system toward Q0347--3819 (normalized intensities are
shown by dots with $1\sigma$ error bars). The zero radial velocity
is fixed at $z_{{\rm H}_2} = 3.024855$.
Smooth lines are the synthetic \hh spectra calculated for the
mean physical parameters (see text). 
Possible blends with lines from the \zabs = 3.025 system are shown without
indicating redshift, whereas blends with lines from other systems
are indicated by ionic species with redshifts in accord with the
list in Appendix B.
\hh profiles disturbed by the Ly-$\alpha$ forest absorption are
also included.
\label{fig5}}
\end{figure}

\begin{figure}
\figcaption[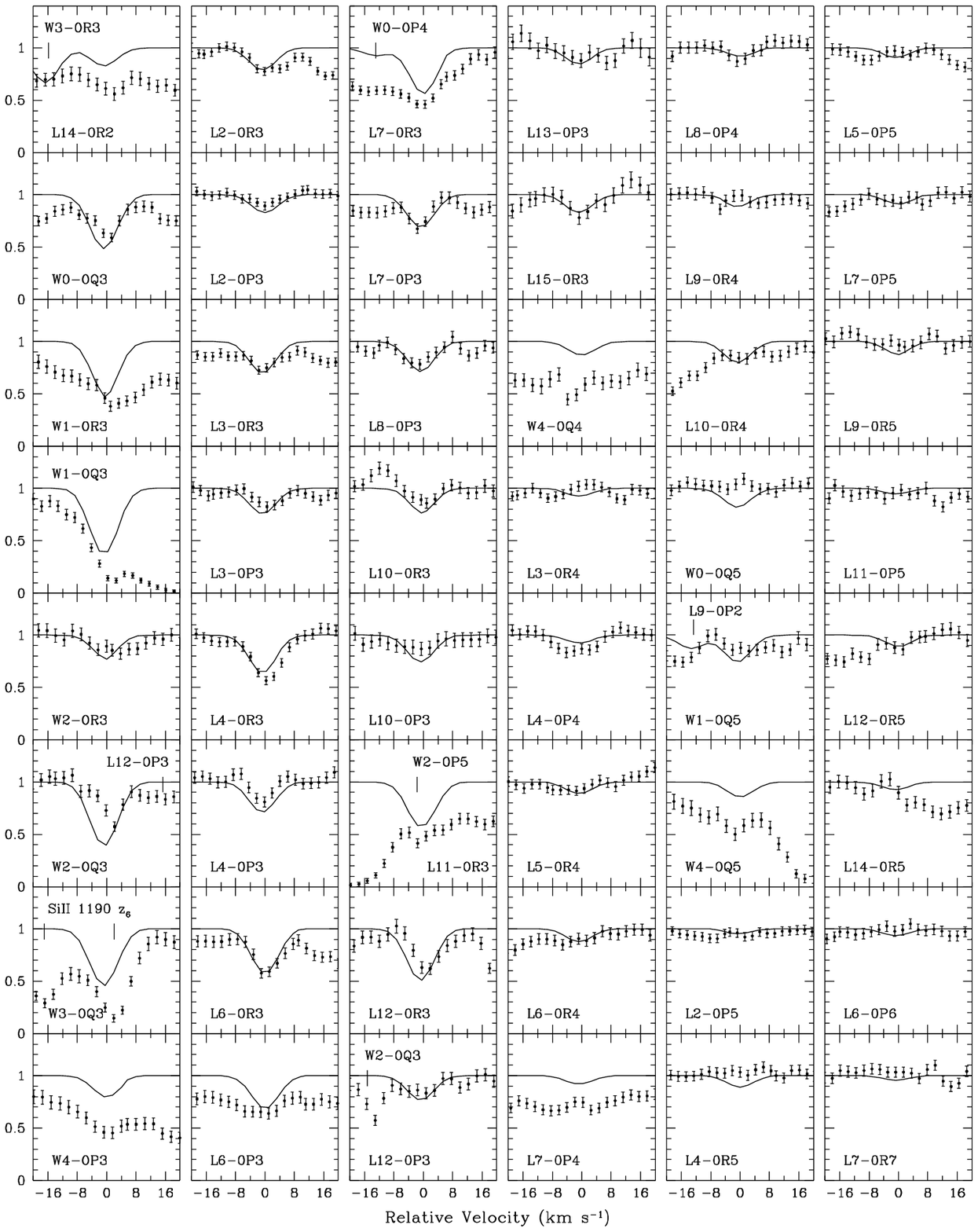]{
Same as Figure~5.
\label{fig6}}
\end{figure}

\begin{figure}
\figcaption[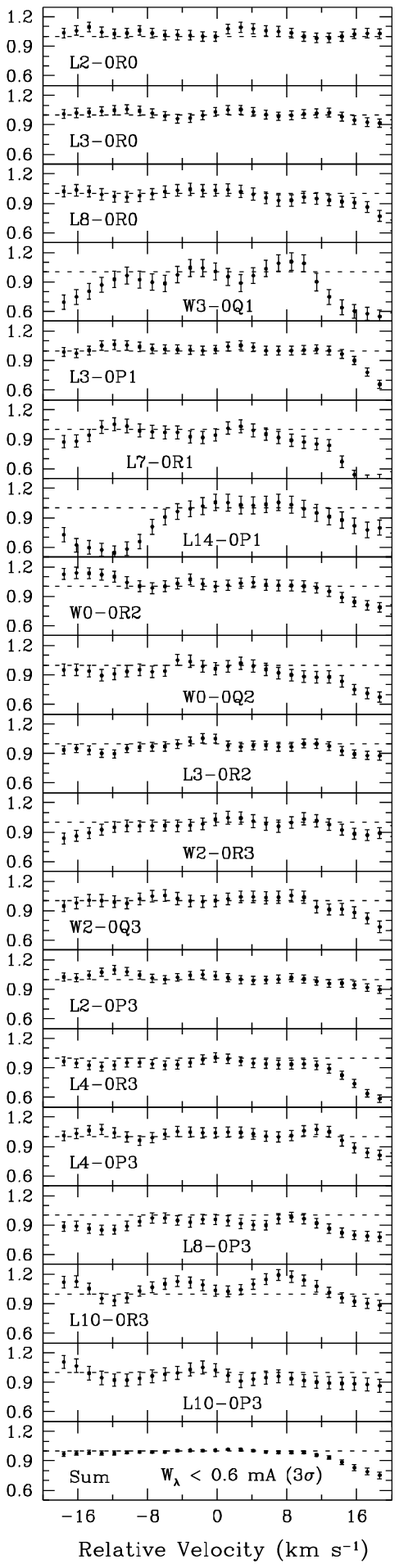]{
Spectral regions including the \hh transitions of the Lyman
and Werner bands (labeled in panels) in the $z = 3.024637$ absorber
in the direction of Q0347--3819. The lowest panel presents the
result of a stacking. The $3\sigma$ upper limit of 0.6 m\AA\, was
calculated for a 12 km~s$^{-1}$ window centered at $v = 0$ km~s$^{-1}$.
\label{fig7}}
\end{figure}

\begin{figure}
\figcaption[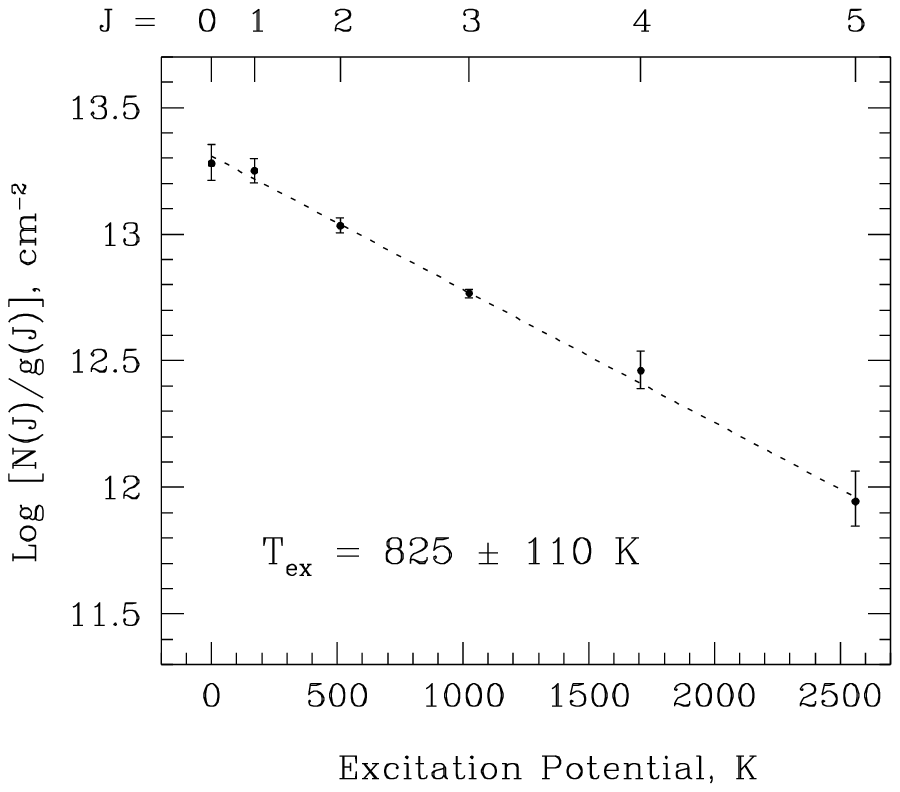]{
Population of the low rotational levels of \hh in
the \zabs = 3.025 damped Ly$\alpha$ system toward
Q0347--3819. Error bars for $1\sigma$ deviations
are shown. The negative inverse slope of the best-fit
straight dotted line indicates the excitation temperature
of about 800~K.
\label{fig8}}
\end{figure}

\begin{figure}
\figcaption[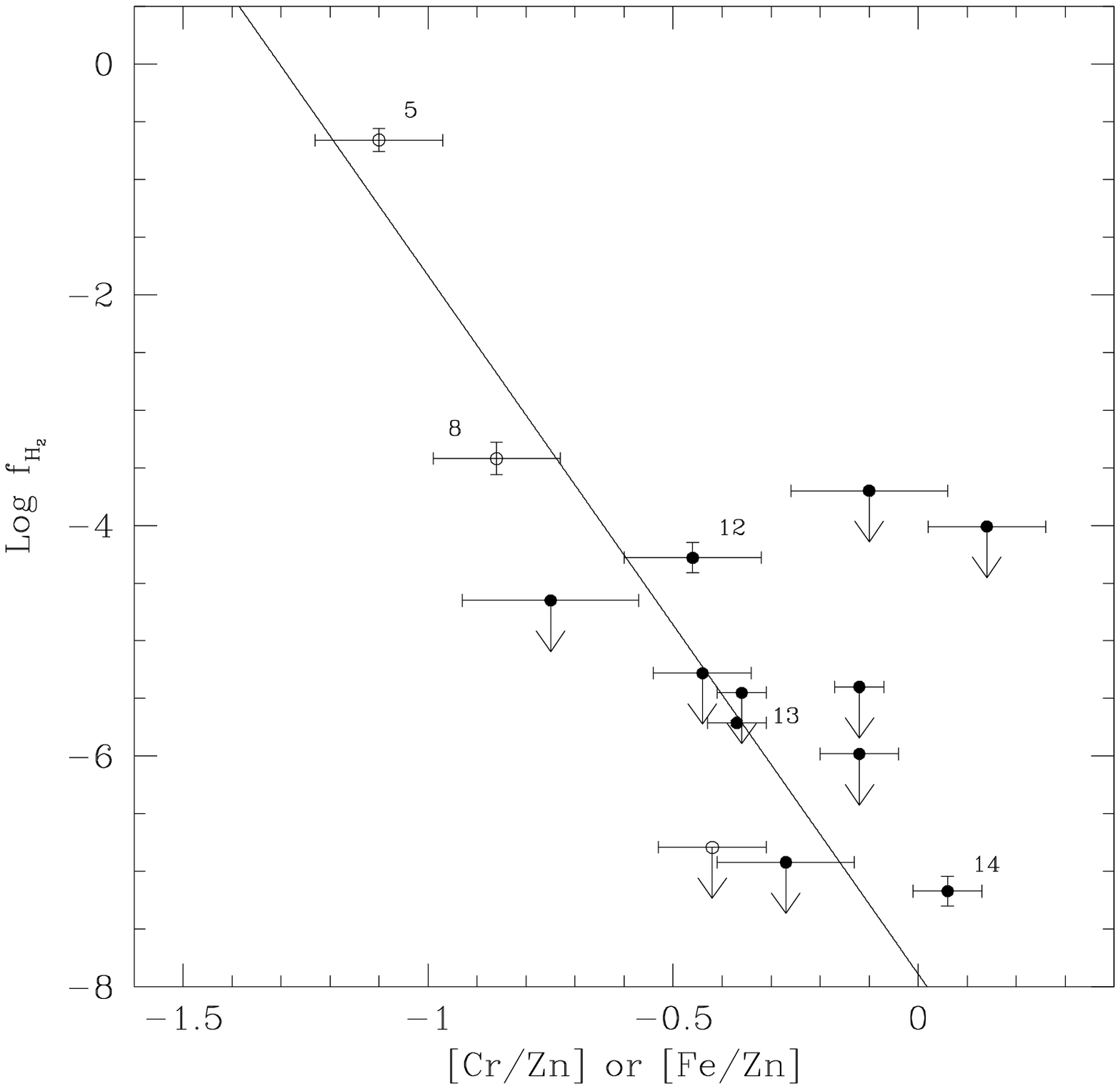]{
Relation between H$_2$ fractional abundance 
$f_{{\rm H}_2}$ plotted on a logarithmic scale 
and relative heavy element depletion in DLAs. 
The solid line corresponds to the linear law: 
$\log f_{{\rm H}_2} \simeq -6.1\,{\rm [Cr/Zn]} - 7.9$, 
obtained for the five measurements of $f_{{\rm H}_2}$ 
we know (DLAs \#~5,~8,~12,~13 and 14). 
The filled circles correspond to [Cr/Zn] and the open ones to 
[Fe/Zn]. See Table~2 for references.
\label{fig9}}
\end{figure}

\begin{figure}
\figcaption[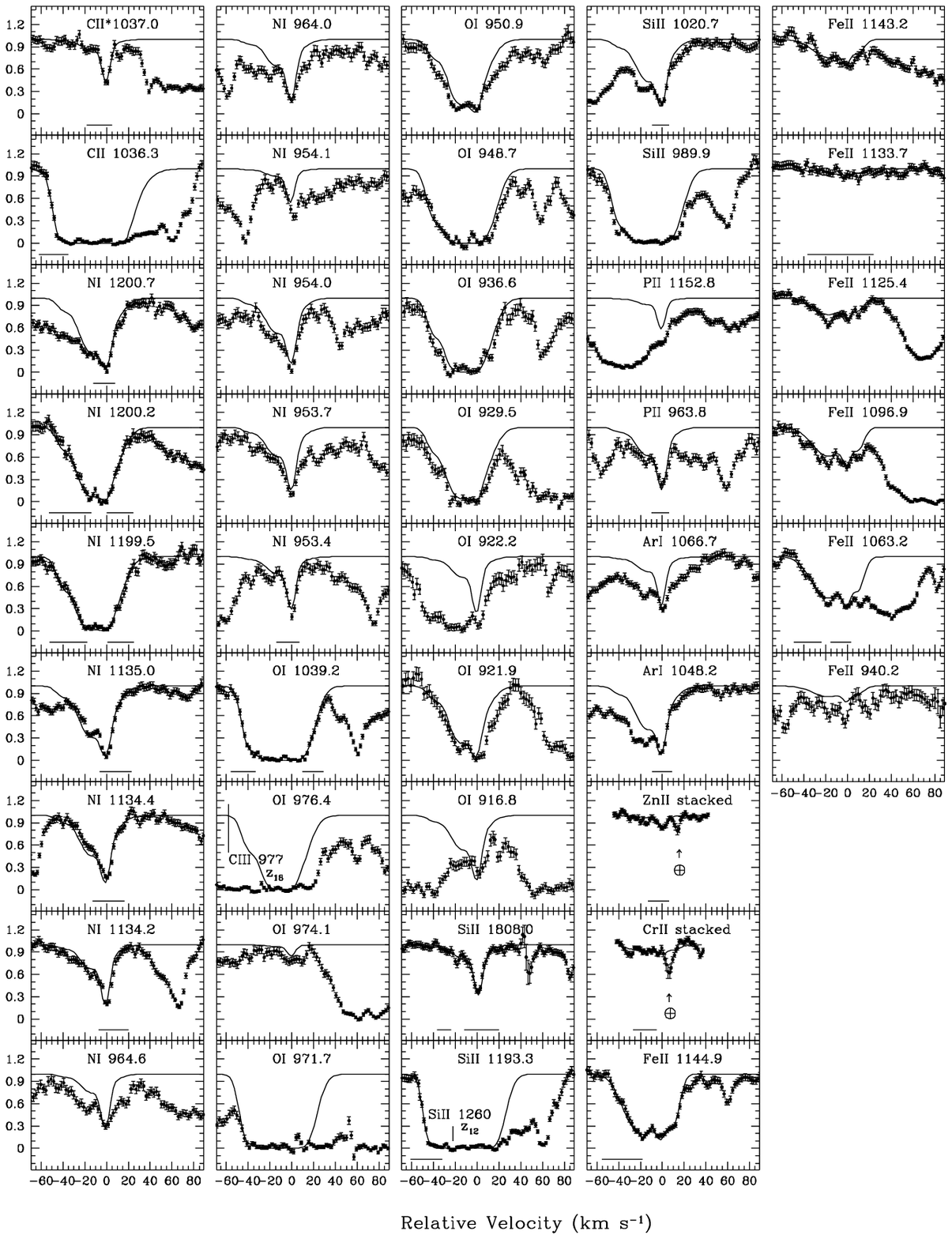]{
Metal absorption features associated with the \zabs = 3.025 DLA
toward Q0347--3819 (normalized intensities are
shown by dots with $1\sigma$ error bars). The zero radial velocity
is fixed at \zabs = 3.024855.
The smooth lines are the synthetic spectra 
convolved with the instrumental resolution of FWHM = 7.0 km~s$^{-1}$.
Two groups of lines 
(\protect\ion{H}{1}, \protect\ion{C}{2}, \protect\ion{C}{2}$^\ast$, \protect\ion{N}{1},
\protect\ion{O}{1}, \protect\ion{Si}{2}, \protect\ion{P}{2}, \protect\ion{Ar}{1}, \protect\ion{Zn}{2},
and
\protect\ion{Fe}{2}, \protect\ion{Cr}{2}) 
are fitted simultaneously
to the observed profiles or to their portions 
marked by horizontal lines which
also indicate pixels involved in the optimization procedure
(the corresponding hydrogen profiles are shown in Figures~11 and 12).
For \protect\ion{Zn}{2} and \protect\ion{Cr}{2} composite stacked spectra are shown. 
Two absorption features at $\Delta v \simeq 17$ km~s$^{-1}$
(\protect\ion{Zn}{2}) and $\Delta v \simeq 6$ km~s$^{-1}$
(\protect\ion{Cr}{2}) are telluric lines.
Possible blends with lines from other systems
are indicated by ionic species with redshifts in accord with the
list in Appendix B.
\label{fig10} }
\end{figure}

\begin{figure}
\figcaption[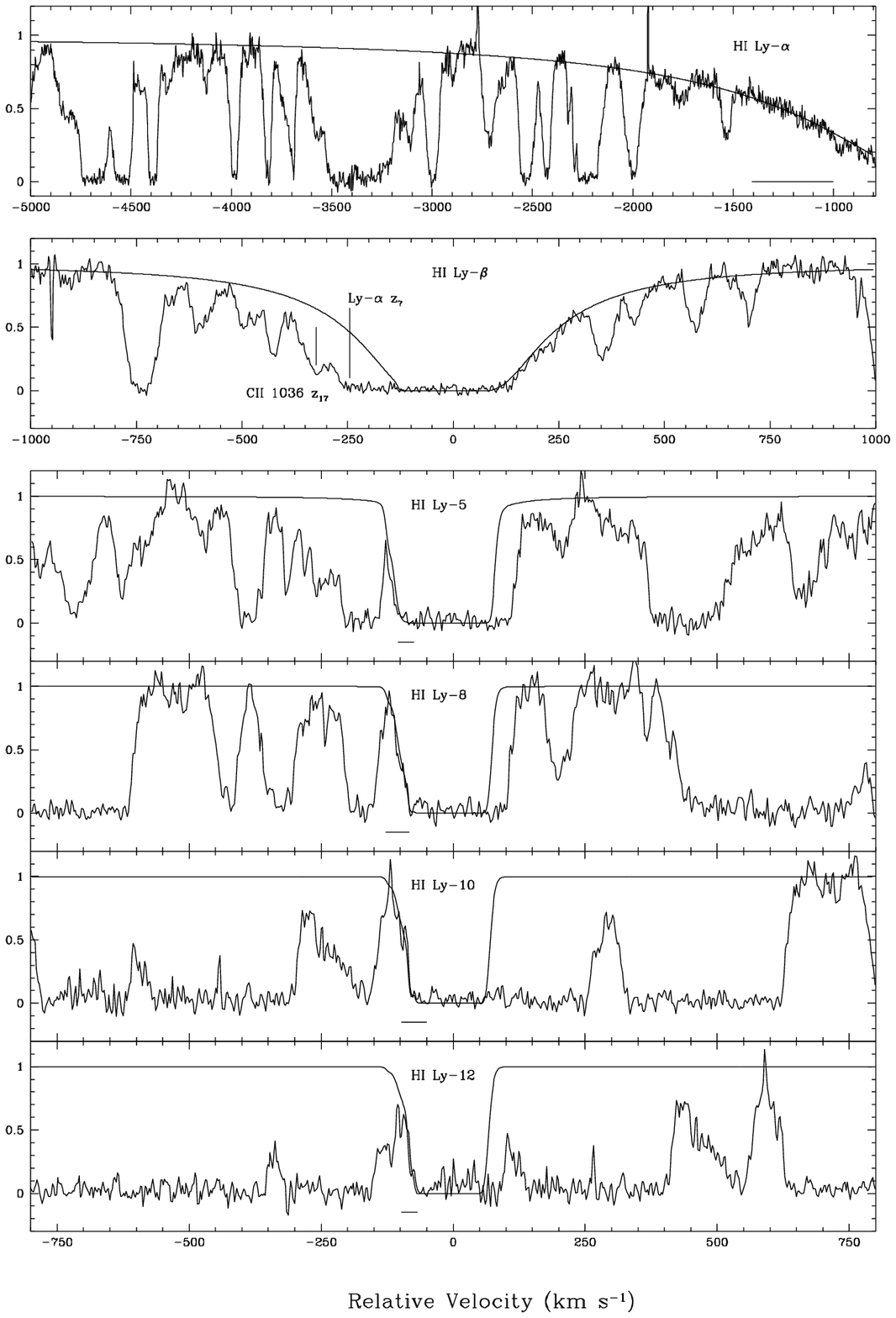]{
Same as Figure~10 but for the hydrogen and deuterium lines.
For the Ly-$\alpha$ line only a portion of its blue wing free of the
forest absorptions is shown.
The blue wing of Ly-$\beta$ is contaminated by the forest absorption. 
The deuterium absorption 
seen in Ly-8, Ly-10 and Ly-12 is shown zoomed in Figure~12.
The corresponding D/H ratio is equal to $3.75\times10^{-5}$.
\label{fig11} }
\end{figure}

\begin{figure}
\figcaption[fig11.ps]{
Same as Figure~11 to show details of the H+D (thick lines)
and D (thin lines) profiles.
\label{fig12} }
\end{figure}

\begin{figure}
\figcaption[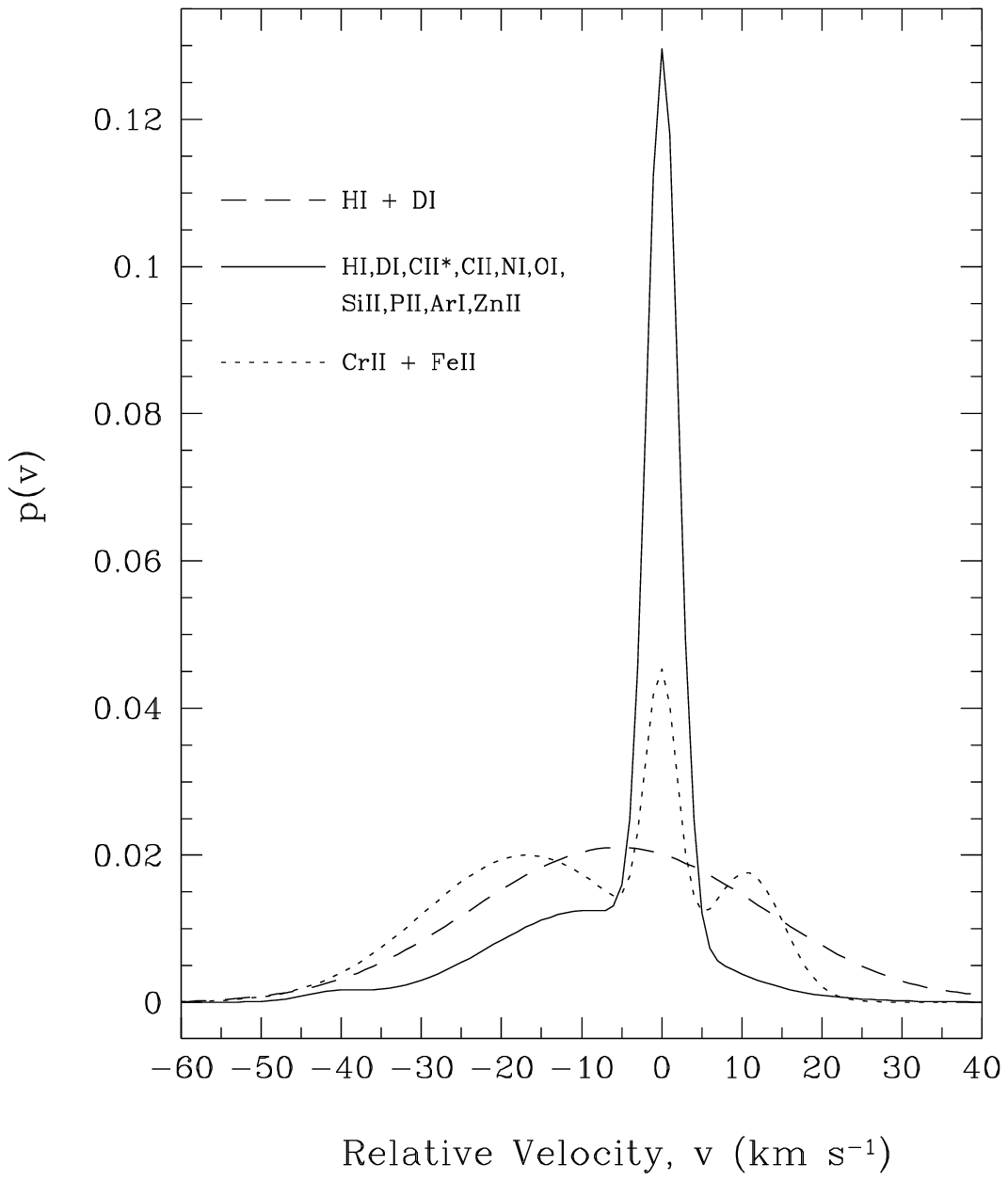]{
Radial velocity distributions $p(v)$ for the models
adopted in the present paper
(solid and dotted curves) and, for comparison, $p(v)$
from DDM (dashed curve).
The difference in shapes of $p(v)$ leads to the difference in the
estimations of D/H: $(2.24\pm0.67)\times10^{-5}$ (DDM) and
$(3.75\pm0.25)\times10^{-5}$ (see Figure~14).    
The intense and narrow spike at $\Delta v \simeq 0$ km~s$^{-1}$
was revealed from the analysis of the \hh lines. 
\label{fig13} }
\end{figure}

\begin{figure}
\figcaption[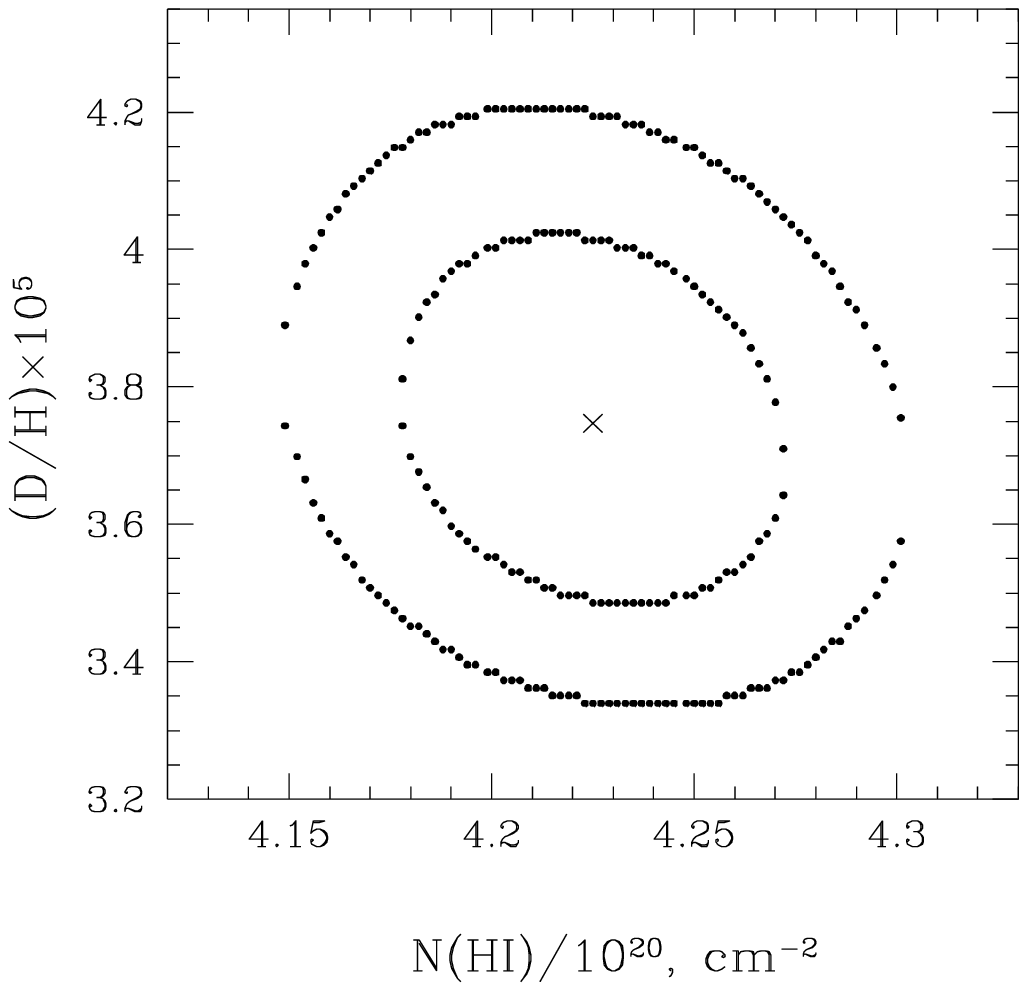]{
Confidence regions in the  `\protect\ion{H}{1} -- D/H' plane calculated from
the simultaneous fit of the \protect\ion{H}{1} and metal lines shown in
Figures~10 and 11 when the other best-fitting parameters of the adopted model
are fixed.  
The contours represent 68.3\% (inner) and 95.4\% (outer) confidence
levels. The cross marks the point of maximum likelihood for the
adopted model ($\chi^2_{\rm min} = 1.22, \nu = 545$).
\label{fig14} }
\end{figure}

\begin{figure}
\figcaption[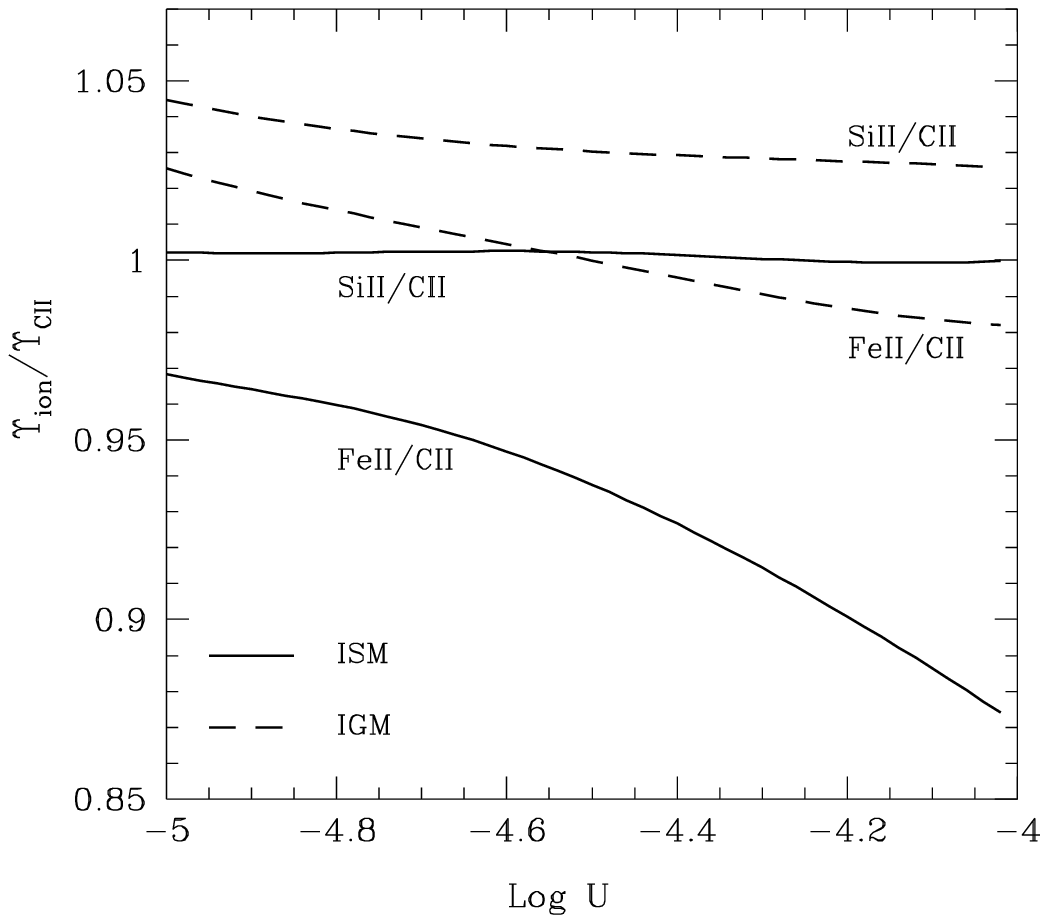]{
Fractional ionization ratios $\Upsilon_{\rm Si\,II}/\Upsilon_{\rm C\,II}$
and $\Upsilon_{\rm Fe\,II}/\Upsilon_{\rm C\,II}$ versus the ionization
parameter $U$ in the case of a soft, starlight background (solid curves)
and a hard, QSO-dominated ionizing field (dashed curves).
Similarity of the line profiles of a pair of ions requires the
fractional ionization ratio to be constant for different $U$.
\label{fig15} }
\end{figure}

\begin{figure}
\figcaption[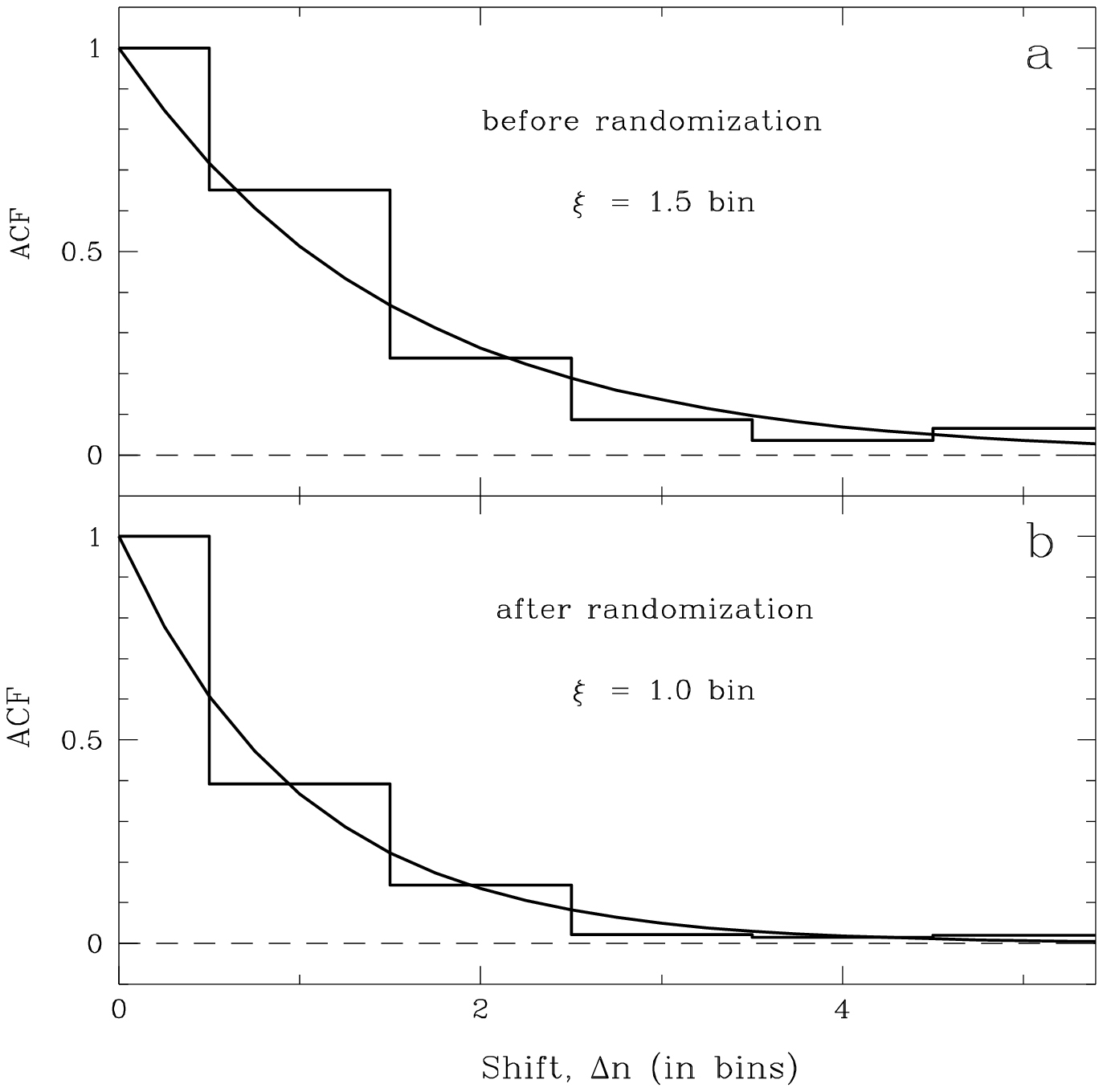]{
Autocorrelation functions (ACF) of the intensity fluctuations
in the continuum between $\lambda\, 6790$ \AA\, and 
$\lambda\, 6820$ \AA\,
in the UVES spectrum of Q0347--3819 before and after randomization
(histograms in panels {\bf a} and {\bf b}, respectively).
The smooth curves show the ACFs of the first order autoregression
process with correlation lengths indicated by $\xi$ (see text for
more details).
\label{fig16} }
\end{figure}

\clearpage

\begin{deluxetable}{clcccccc}
\tabletypesize{\scriptsize}
\tablecaption{H$_2$ {\sc Lines at \zabs = 3.025 toward Q0347--3819 
} 
\label{tbl-1}}
\tablewidth{0pt}
\tablehead{
\colhead{J} & 
\colhead{Line} & \colhead{$\langle\lambda\rangle$, \AA} & 
\colhead{$\sigma_{\langle\lambda\rangle}$, \AA} &
\colhead{$W_\lambda$, m\AA} & \colhead{$\sigma_{W_\lambda}$, m\AA}  
& \colhead{$z$} & \colhead{$\sigma_z$} 
}
\startdata
0 & L3-0R$^a$ & 4277.96 & 0.09 & 6.7 & 4.4 & 3.02486 & 0.00009  \\
  & L7-0R$^{a,b}$ & 4076.44 & 0.06 & 25.8 & 5.7 & 3.02485 & 0.00006 \\
1 & W2-0Q$^b$ & 3888.38 & 0.03 & 83.9 & 4.4 & 3.02487 & 0.00003 \\
  & W3-0Q$^b$ & 3813.22 & 0.05 & 55.7 & 7.2 & 3.02484 & 0.00005 \\
  & L1-0P$^b$ & 4403.42 & 0.04 & 25.4 & 4.1 & 3.02486 & 0.00004  \\
  & L3-0R$^b$ & 4280.27 & 0.03 & 65.6 &  3.8 & 3.02485 & 0.00002 \\
  & L3-0P$^{a,b}$ & 4284.89 & 0.04 & 24.2 &  4.2 & 3.02486 & 0.00004 \\
  & L7-0R$^b$ & 4078.95 & 0.03 & 76.4 & 4.8 & 3.02485 & 0.00003 \\
  & L10-0P$^b$ & 3955.77 & 0.04 & 46.8 & 5.0 & 3.02484 & 0.00004 \\
  & L14-0P$^{a,b}$ & 3813.61 & 0.09 & 24.6 & 7.2 & 3.02484 & 0.00009  \\
2 & W0-0R$^{a,b}$ & 4061.19 & 0.05 & 34.4 & 5.5 & 3.02488 & 0.00005 \\
  & W0-0Q$^{a,b}$ & 4068.87 & 0.04 & 50.7 & 6.7 & 3.02484 & 0.00004 \\
  & W0-0P$^b$ & 4073.85 & 0.09 & 10.3 & 4.1 & 3.02488 & 0.00009 \\
  & W1-0Q$^b$ & 3976.45 & 0.04 & 36.6 & 4.4 & 3.02487 & 0.00004 \\
  & L3-0R$^{a,b}$ & 4286.44 & 0.06 & 14.9 & 4.7 & 3.02485 & 0.00006 \\
  & L3-0P$^b$ & 4294.12 & 0.07 & 12.6 &  3.9 & 3.02485 & 0.00007 \\
  & L4-0R$^{a,b}$ & 4232.14 & 0.05 & 24.9 & 5.0 & 3.02487 & 0.00005 \\
  & L5-0R$^{a,b}$ & 4180.57 & 0.04 & 32.2 & 4.4 & 3.02485 & 0.00004 \\
3 & W0-0Q$^b$ & 4075.88 & 0.03 & 65.7 & 5.0 & 3.02485 & 0.00003 \\
  & L2-0R$^b$ & 4353.77 & 0.03 & 36.1 &  4.6 & 3.02488 & 0.00003 \\
  & L2-0P$^a$ & 4365.17 & 0.07 & 8.6 & 5.7 & 3.02481 & 0.00007 \\
  & L3-0R$^b$ & 4296.43 & 0.03 & 51.4 & 4.6 & 3.02484 & 0.00003 \\
  & L3-0P$^{a,b}$ & 4307.18 & 0.05 & 22.0 & 4.5 & 3.02487 & 0.00005 \\
  & L4-0R$^b$ & 4252.15 & 0.03 & 61.1 & 5.4 & 3.02486 & 0.00003 \\
  & L4-0P$^a$ & 4252.14 & 0.09 & 8.4 &  6.0 & 3.02486 & 0.00008 \\
  & L6-0R$^b$ & 4141.54 & 0.03 & 60.3 & 4.6 & 3.02487 & 0.00003 \\
  & L7-0P$^b$ & 4103.35 & 0.05 & 32.2 &  4.0 & 3.02484 & 0.00005 \\
  & L8-0P$^{a,b}$ & 4058.61 & 0.05 & 30.1 & 5.5 & 3.02484 & 0.00005 \\
  & L12-0R$^{a,b}$ & 3894.77 & 0.04 & 49.0 & 6.2 & 3.02488 & 0.00005 \\
  & L15-0R$^{a,b}$ & 3795.31 & 0.08 & 20.8 & 5.0 & 3.02486 & 0.00008 \\
4 & L4-0P$^{a,b}$ & 4268.64 & 0.05 & 24.1 & 5.9 & 3.02482 & 0.00005 \\
  & L8-0P$^a$ & 4074.21 & 0.09 & 8.6 & 5.0 & 3.02484 & 0.00009 \\
5 & W1-0Q$^{a,b}$ & 4004.44 & 0.06 & 25.7 & 5.4 & 3.02489 & 0.00006 \\
  & L7-0P$^{a,b}$ & 4138.55 & 0.10 &  8.5 & 4.7 & 3.02484 & 0.00010 \\
  & L12-0R$^{a,b}$ & 3923.76 & 0.08 & 13.9 & 4.4 & 3.02484 & 0.00008 \\
 \enddata
\tablecomments{$^a$\,Lines are used in the Voigt-fitting procedure
to estimate the broadening $b$-parameter. $^b$\,Lines are
used in Figure~3.
}
 \end{deluxetable}

\clearpage

\begin{deluxetable}{clcccccc}
\tabletypesize{\scriptsize}
\tablecaption{{\sc 
Molecular hydrogen and metal abundance measurements in DLAs 
} \label{tbl-2}}
\tablewidth{0pt}
\tablehead{
\colhead{\#} & 
\colhead{QSO} & \colhead{\zabs} & 
\colhead{[Cr/Zn]} &
\colhead{[Fe/Zn]} & \colhead{$\log N$(H\,{\sc i})}  
& \colhead{$\log N$(H$_2$)} & \colhead{$\log f_{{\rm H}_2}$} 
}
\startdata
1&Q1328+307&0.692&$-0.44\pm0.10^{a}$&$-$ & $21.28\pm0.10^{a}$
&$< 15.70^{b}$& $<-5.28$ \\
2&Q0454+039&0.860&$0.14\pm0.12^{c}$&$0.01\pm0.12^{c}$
&$20.69\pm0.06^{c}$&$< 16.38^{b}$&$<-4.01$ \\
3&Q0935+417&1.373&$-0.10\pm0.16^{d}$&$-0.27\pm0.15^{d}$
&$20.40\pm0.09^{d}$&$<16.40^{b}$&$<-3.70$ \\
4&Q1157+014&1.944&$-$&$-0.42\pm0.11^{e}$&$21.80\pm0.10^{f}$
&$< 14.71^{e}$&$<-6.79$ \\          
5&Q0013$-$004&1.973&$<-0.95^{g}$&$-1.10\pm0.13^{g,h}$
&$20.70\pm0.05^{i}$&$19.84\pm0.10^{i}$&$-0.66\pm0.10$ \\
6&Q0458$-$020&2.040&$-0.36\pm0.05^{j}$&$-0.46\pm0.07^{j}$
&$21.65\pm0.09^{g}$&$< 15.90^{b}$&$<-5.45$ \\
7&Q0100+130&2.309&$-0.12\pm0.05^{j}$&$-0.22\pm0.06^{j}$
&$21.40\pm0.05^{k}$&$< 15.70^{b}$&$< -5.40$ \\           
8&Q1232+082&2.338&$-$&$-0.86\pm 0.13^{l,m}$&$20.90\pm0.10^{m}$&
$17.18\pm0.10^{m}$&$-3.42\pm0.14$ \\
9&Q0841+129&2.374&$-0.12\pm0.08^{e,j}$&$-0.29\pm0.08^{e}$
&$20.95\pm0.09^{j}$&$<14.67^{e}$&$< -5.98$ \\
10&Q0112+029&2.423&$-0.75\pm0.18^{g}$&$-$&$20.95\pm0.10^{g}$
&$< 16.00^{b}$&$<-4.65$ \\
11&Q1223+178&2.466&$-0.27\pm0.14^{g,e}$&$-0.21\pm0.11^{e}$
&$21.52\pm0.10^{e}$&$< 14.30^{e}$&$< -6.92$ \\
12&Q0528$-$250&2.811&$-0.46\pm0.14^{n}$&$-0.46\pm0.12^{n,o}$
&$21.35\pm0.10^{u,p}$&$16.77\pm0.09^{q,o}$& $-4.28\pm 0.13$ \\
13& Q0347$-$382&3.025&$-0.37\pm0.06^{r}$&$-0.19\pm0.06^{r}$
&$20.626\pm0.005^{r}$
&$14.613\pm0.022^{r}$&$-5.71\pm0.02$ \\
14&Q0000$-$262&3.390&$0.06\pm0.07^{t}$&$0.03\pm0.07^{t}$
&$21.41\pm0.08^{j}$& 
$13.94\pm 0.06^{v,w}$&$-7.17\pm 0.13$ \\
\enddata
\tablecomments{Column densities $N$(H\,{\sc i})
and $N$(H$_2$) are given in cm$^{-2}$.\\ References: 
$^a$~Meyer \& York (1992); 
$^b$~Ge \& Bechtold (1999); 
$^c$~Pettini et al. (2000); 
$^d$~Meyer et al. (1995); $^e$ Petitjean et al. (2000); 
$^f$~Wolfe \& Briggs (1981); $^g$ Pettini et al. (1994); 
$^h$~Centurion et al. (2000); $^i$ Ge \& Bechtold (1997); 
$^j$~Prochaska \& Wolfe (1999); 
$^k$~Wolfe et al. (1994); $^l$ Ge et al. (2001); 
$^m$~Srianand et al. (2001); $^n$ Lu et al. (1996); 
$^o$~Srianand \& Petitjean (1998); 
$^p$~M\o ller \& Warren (1993); 
$^q$~Levshakov \& Varshalovich (1985); 
$^r$~this paper; 
$^t$~Molaro et al. (2000); 
$^u$~Levshakov \& Foltz (1988);
$^v$~Levshakov et al. (2000); 
$^w$~Levshakov et al. (2001) }

\end{deluxetable}

\clearpage

\begin{deluxetable}{lcccr}
\tabletypesize{\scriptsize}
\tablecaption{{\sc 
Column densities and relative abundances 
in the H$_2$-bearing cloud at \zabs = 3.025
toward Q0347--3819
} \label{tbl-3}}
\tablehead{
\colhead{Ion} & 
\colhead{log N, cm$^{-2}$} & 
\colhead{(X/H)$^b_\odot$} & 
\colhead{[X/H]} &
\colhead{[X/Zn]} 
}
\startdata
H\,{\sc i}&$20.402\pm0.006^a$& & & \\
C\,{\sc i} & $< 11.6^{\dagger,\ddagger}$ & & & \\
C\,{\sc ii}$^\ast$&$13.354\pm0.023$& & & \\
C\,{\sc ii}&$15.703\pm0.024$&$-3.48\pm0.06$&$-1.22\pm0.06$&$0.29\pm0.08$ \\
N\,{\sc i}&$14.664\pm0.007$&$-4.08\pm0.06$&$-1.66\pm0.06$&$-0.15\pm0.08$ \\
O\,{\sc i}$^\ast$ & $< 11.9^{\dagger,\ddagger}$ & & & \\
O\,{\sc i}&$16.410\pm0.010$&$-3.17\pm0.06$&$-0.82\pm0.06$&$0.69\pm0.08$ \\
Si\,{\sc ii}$^\ast$ & $< 12.4^{\dagger,\ddagger}$ & & & \\
Si\,{\sc ii}&$15.015\pm0.013$&$-4.44\pm0.05$&$-0.95\pm0.02$&$0.56\pm0.06$ \\
P\,{\sc ii}&$12.779\pm0.032$&$-6.44\pm0.06$&$-1.18\pm0.07$&$0.32\pm0.09$ \\
Ar\,{\sc i}&$13.760\pm0.020$&$-5.48\pm0.04^c$&$-1.16\pm0.05$&$0.35\pm0.07$ \\
Cr\,{\sc ii}&$11.698\pm0.046$&$-6.31\pm0.01$&$-2.39\pm0.05$&$-0.89\pm0.07$ \\
Fe\,{\sc ii}&$13.690\pm0.032$&$-4.50\pm0.01$&$-2.21\pm0.03$&$-0.71\pm0.06$ \\
Zn\,{\sc ii}&$11.565\pm0.037$&$-7.33\pm0.04$&$-1.51\pm0.05$& 
\enddata
\tablecomments{
$^a$\,Indicated are the internal errors.
$^\dagger$\,$1\sigma$ level; $^\ddagger$\,based on the Keck/HIRES
data obtained by Prochaska \& Wolfe (1999). 
References~:\, 
$^b$ Grevesse \& Sauval (1998);
$^c$ from Sofia \& Jenkins (1998).
}

\end{deluxetable}

\clearpage

\begin{deluxetable}{lccccrrr}
\tabletypesize{\scriptsize}
\tablecaption{{\sc 
Total column densities and relative abundances at \zabs = 3.025
toward Q0347--3819$^\dagger$
} \label{tbl-4}}
\tablewidth{0pt}
\tablehead{
\colhead{Ion} & 
\colhead{log N, cm$^{-2}$} & 
\colhead{(X/H)$^d_\odot$} & 
\colhead{[X/H]} &
\colhead{[X/Zn]} & 
\colhead{[X/H]$^b$} & 
\colhead{[X/H]$^c$} &
\colhead{[X/H]$^e$} 
}
\startdata
H\,{\sc i} & $20.626\pm0.005^x$ & & & & & &  \\
C\,{\sc ii}&$15.929\pm0.024$& $-3.48\pm0.06$&$-1.22\pm0.06$&
$0.29\pm0.08$ & & $>-2.4$ & $>-2.2$ \\
C\,{\sc ii}$^\ast$ & $13.581\pm0.024$ & & & & & &  \\
N\,{\sc i}&$14.890\pm0.007$&$-4.08\pm0.06$ & $-1.66\pm0.06$ & 
$-0.15\pm0.08$ & $-2.1$ & $-2.5$ &  \\
O\,{\sc i}&$16.636\pm0.011$&$-3.17\pm0.06$ & $-0.82\pm0.06$ &
$0.68\pm0.08$ & $>-1.5$ & $>-2.4$ & $>-2.2$ \\
Si\,{\sc ii}&$15.240\pm0.013$ & $-4.44\pm0.05$ & $-0.95\pm0.02$&
$0.56\pm0.06$& $-1.2$ & $>-1.9$ & $-1.4$ \\
P\,{\sc ii}&$13.005\pm0.031$&$-6.44\pm0.06$ & $-1.18\pm0.07$ &
$0.32\pm0.09$ & & & \\
Ar\,{\sc i}&$13.986\pm0.020$&$-5.48\pm0.04^a$&$-1.16\pm0.04$&
$0.34\pm0.07$& & & \\
Cr\,{\sc ii}&$12.441\pm0.035$ & $-6.31\pm0.01$ & $-1.88\pm0.04$& 
$-0.37\pm0.06$ & & & \\
Fe\,{\sc ii}&$14.432\pm0.009$&$-4.50\pm0.01$&$-1.69\pm0.01$& 
$-0.19\pm0.06$ & $-1.9$ & $>-2.0$ &$-1.7$\\
Zn\,{\sc ii}&$11.792\pm0.036$&$-7.33\pm0.04$&$-1.50\pm0.05$& &$<-0.8$ & $<-1.2$ &  
\enddata
\tablecomments{$^\dagger$\,Listed values were measured in the range
$-50$ km~s$^{-1}$ $ \leq v \leq 20$ km~s$^{-1}$ (see Figure~10).
$^x$\,Indicated are the internal errors.
References~:\, 
$^a$ from Sofia \& Jenkins (1998);
$^b$ Centuri\'on et al. (1998);
$^c$ Ledoux et al. (1998);
$^d$ Grevesse \& Sauval (1998);
$^e$ Prochaska \& Wolfe (1999)
}
\end{deluxetable}

\clearpage

\begin{deluxetable}{clcccc}
\tabletypesize{\scriptsize}
\tablecaption{H$_2$ {\sc Transition probabilities and excitation rate
coefficients for $\Delta${\rm J} = 2 transitions between the first
five rotational states}
\label{tbl-5}}
\tablewidth{0pt}
\tablehead{
\colhead{J} & 
\colhead{$A_{{\rm J+2}\rightarrow{\rm J}}\,
\left(\frac{1}{\rm s}\right)^a$} & 
\colhead{$q^{T=400{\rm K}}_{{\rm J}\rightarrow{\rm J+2}}\,
\left(\frac{{\rm cm}^3}{\rm s}\right)^b$} & 
\colhead{$q^{T=1000{\rm K}}_{{\rm J}\rightarrow{\rm J+2}}\,
\left(\frac{{\rm cm}^3}{\rm s}\right)^b$} & 
\colhead{$q^{T=1000{\rm K}}_{{\rm J}\rightarrow{\rm J+2}}\,
\left(\frac{{\rm cm}^3}{\rm s}\right)^c$} & 
\colhead{$q^{T=1000{\rm K}}_{{\rm J}\rightarrow{\rm J+2}}\,
\left(\frac{{\rm cm}^3}{\rm s}\right)^d$}  
}
\startdata
0 & 2.94(-11) & 1.14(-11) & 6.44(-11) & 4.82(-11) & 1.27(-10) \\
1 & 4.76(-10) & 1.79(-12) & 1.60(-11) & $\ldots$  & 4.07(-11) \\
2 & 2.76(-9)  & 5.16(-13) & 7.35(-12) & 6.70(-12) & 1.78(-11) \\
3 & 9.84(-9)  & 1.76(-13) & 4.02(-12) & $\ldots$  & 8.33(-12)
\enddata
\tablecomments{The negative numbers in parenthesis
are the powets of ten. References~:\, $^a$~Turner et al. (1977); 
$^b$~this paper;
$^c$~Elitzur \& Watson (1978);
$^d$~Nishimura (1968)
}
\end{deluxetable}

\clearpage

\begin{deluxetable}{lcccc}
\tabletypesize{\scriptsize}
\tablecaption{{\sc Comparison of the mean values and
the standard deviations estimated
by the Monte Carlo (MC) and $\Delta\chi^2$
methods in case of correlated measurements
}
\label{tbl-6}}
\tablewidth{0pt}
\tablehead{
\colhead{parameter, $x$} &
\colhead{$\langle x \rangle_{\rm MC}$} & \colhead{$\sigma_{\rm MC}$} &
\colhead{$\langle x \rangle_{\Delta\chi^2}$} & 
\colhead{$\sigma_{\Delta\chi^2}$} 
}
\startdata
$b^\ddagger$, km~s$^{-1}$ &  3.04  &   0.29  &   2.80 &  0.45 \\
$N$(J=0), cm$^{-2}$ & $1.90^\dagger_{13}$ & $0.29_{13}$ &
$1.90_{13}$ & $0.35_{13}$ \\
$N$(J=1), cm$^{-2}$ & $1.84_{14}$ & $0.13_{14}$ & $1.60_{14}$ &
$0.18_{14}$ \\
$N$(J=2), cm$^{-2}$ & $5.39_{13}$ & $0.28_{13}$ & $5.40_{13}$ &
$0.35_{13}$ \\
$N$(J=3), cm$^{-2}$ & $1.24_{14}$ & $0.07_{14}$ & $1.22_{14}$ &
$0.05_{14}$ \\
$N$(J=4), cm$^{-2}$ & $2.35_{13}$ & $0.44_{13}$ & $2.60_{13}$ &
$0.43_{13}$ \\
$N$(J=5), cm$^{-2}$ & $2.90_{13}$ & $0.53_{13}$ & $2.90_{13}$ &
$0.70_{13}$ 
\enddata
\tablecomments{$^\ddagger$\,The \hh column densities at
different J levels and the corresponding broadening $b$
parameter are listed in column (1).\,
$^\dagger$\,Values like `$a_t$' mean $a\times10^t$. 
}
\end{deluxetable}

\end{document}